\documentclass{aastex}
\usepackage{natbib}
\usepackage{graphicx}
\usepackage{amsmath}
\usepackage{subfigure}
\usepackage{rotating}

\shorttitle{High Frequency Periodicity Studies in SNO}
\shortauthors{B. Aharmin \emph{et al.} (SNO Collaboration)}

\begin{document}
\title{Searches for High Frequency Variations \\
in the $^8$B Solar Neutrino Flux at the Sudbury Neutrino Observatory}

\newcommand{\alta}{Department of Physics, University of
Alberta, Edmonton, Alberta, T6G 2R3, Canada}
\newcommand{\ubc}{Department of Physics and Astronomy, University of
British Columbia, Vancouver, BC V6T 1Z1, Canada}
\newcommand{\bnl}{Chemistry Department, Brookhaven National
Laboratory,  Upton, NY 11973-5000}
\newcommand{\carleton}{Ottawa-Carleton Institute for Physics, Department of
Physics, Carleton University, Ottawa, Ontario K1S 5B6, Canada}
\newcommand{\casa}{Center for Astrophysics and Space Astronomy, University of Colorado, 
Boulder, CO}
\newcommand{\uog}{Physics Department, University of Guelph,
Guelph, Ontario N1G 2W1, Canada}
\newcommand{\lu}{Department of Physics and Astronomy, Laurentian
University, Sudbury, Ontario P3E 2C6, Canada}
\newcommand{\lbnl}{Institute for Nuclear and Particle Astrophysics and
Nuclear Science Division, Lawrence Berkeley National Laboratory, Berkeley, CA
94720}
\newcommand{\lbla}{ Lawrence Berkeley National Laboratory, Berkeley, CA}
\newcommand{\lanl}{Los Alamos National Laboratory, Los Alamos, NM 87545}
\newcommand{\llnl}{Lawrence Livermore National Laboratory, Livermore, CA}
\newcommand{\lanla}{Los Alamos National Laboratory, Los Alamos, NM 87545}
\newcommand{\oxford}{Department of Physics, University of Oxford,
Denys Wilkinson Building, Keble Road, Oxford OX1 3RH, UK}
\newcommand{\penn}{Department of Physics and Astronomy, University of
Pennsylvania, Philadelphia, PA 19104-6396}
\newcommand{\queens}{Department of Physics, Queen's University,
Kingston, Ontario K7L 3N6, Canada}
\newcommand{\presqueens}{Department of Physics, Queen's University,
Kingston, Ontario, Canada}
\newcommand{\presuw}{Center for Experimental Nuclear Physics and Astrophysics,
and Department of Physics, University of Washington, Seattle, WA} 
\newcommand{\uw}{Center for Experimental Nuclear Physics and Astrophysics,
and Department of Physics, University of Washington, Seattle, WA 98195} 
\newcommand{\uta}{Department of Physics, University of Texas at Austin,
Austin, TX 78712-0264}
\newcommand{\triumf}{TRIUMF, 4004 Wesbrook Mall, Vancouver, BC V6T 2A3,
Canada}
\newcommand{\ralimp}{Rutherford Appleton Laboratory, Chilton, Didcot OX11 0QX,
UK}
\newcommand{\iusb}{Department of Physics and Astronomy, Indiana University,
South Bend, IN}
\newcommand{\fnal}{Fermilab, Batavia, IL}
\newcommand{\uo}{Department of Physics and Astronomy, University of Oregon,
Eugene, OR}
\newcommand{\hu}{Department of Physics, Hiroshima University, Hiroshima,
Japan}
\newcommand{\slac}{Stanford Linear Accelerator Center, Menlo Park, CA}
\newcommand{\homestake}{Sanford Laboratory at Homestake, Lead, SD}
\newcommand{\mac}{Department of Physics, McMaster University, Hamilton, ON}
\newcommand{\doe}{US Department of Energy, Germantown, MD} 
\newcommand{\lund}{Department of Physics, Lund University, Lund, Sweden}
\newcommand{\mpi}{Max-Planck-Institut for Nuclear Physics, Heidelberg,
Germany}
\newcommand{\uom}{Ren\'{e} J.A. L\'{e}vesque Laboratory, Universit\'{e} de
Montr\'{e}al, Montreal, PQ}
\newcommand{\cwru}{Department of Physics, Case Western Reserve University,
Cleveland, OH}
\newcommand{\pnnl}{Pacific Northwest National Laboratory, Richland, WA} 
\newcommand{\uc}{Department of Physics, University of Chicago, Chicago, IL}
\newcommand{\mitt}{Laboratory for Nuclear Science, Massachusetts Institute of
Technology, Cambridge, MA 02139}
\newcommand{\ucsd}{Department of Physics, University of California at San
Diego, La Jolla, CA }
\newcommand{    \lsu    }{Department of Physics and Astronomy, Louisiana State
University, Baton Rouge, LA 70803}
\newcommand{\imp}{Imperial College, London SW7 2AZ, UK}
\newcommand{\presimp}{Imperial College, London, UK}
\newcommand{\uci}{Department of Physics, University of California, Irvine, CA
92717}
\newcommand{\ucia}{Department of Physics, University of California, Irvine,
CA}
\newcommand{\suss}{Department of Physics and Astronomy, University of Sussex,
Brighton  BN1 9QH, UK}
\newcommand{\pressuss}{Department of Physics and Astronomy, University of Sussex,
Brighton, UK}
\newcommand{    \lifep  }{Laborat\'{o}rio de Instrumenta\c{c}\~{a}o e
F\'{\i}sica Experimental de
Part\'{\i}culas, Av. Elias Garcia 14, 1$^{\circ}$, 1000-149 Lisboa, Portugal}
\newcommand{\savannah}{Department of Chemistry and Physics, Armstrong Atlantic 
State University, Savannah, GA}
\newcommand{\hku}{Department of Physics, The University of Hong Kong, Hong
Kong.}
\newcommand{\aecl}{Atomic Energy of Canada, Limited, Chalk River Laboratories,
Chalk River, ON K0J 1J0, Canada}
\newcommand{\nrc}{National Research Council of Canada, Ottawa, ON K1A 0R6,
Canada} 
\newcommand{\princeton}{Department of Physics, Princeton University,
Princeton, NJ 08544}
\newcommand{\birkbeck}{Birkbeck College, University of London, Malet Road,
London WC1E 7HX, UK}
\newcommand{\snoi}{SNOLAB, Sudbury, ON P3Y 1M3, Canada}
\newcommand{\uba}{University of Buenos Aires, Argentina}
\newcommand{\hvd}{Department of Physics, Harvard University, Cambridge, MA}
\newcommand{\mun}{gotta check this one}
\newcommand{\pny}{Goldman Sachs, 85 Broad Street, New York, NY}
\newcommand{\pnv}{Remote Sensing Lab, PO Box 98521, Las Vegas, NV 89193}
\newcommand{\psis}{Paul Schiffer Institute, Villigen, Switzerland}
\newcommand{\liverpool}{Department of Physics, University of Liverpool,
Liverpool, UK}
\newcommand{\uto}{Department of Physics, University of Toronto, Toronto, ON,
Canada}
\newcommand{\uwisc}{Department of Physics, University of Wisconsin, Madison,
WI}
\newcommand{\psu}{Department of Physics, Pennsylvania State University,
     University Park, PA}
\newcommand{\anl}{Deparment of Mathematics and Computer Science, Argonne
     National Laboratory, Lemont, IL}
\newcommand{\cornell}{Department of Physics, Cornell University, Ithaca, NY}
\newcommand{\tufts}{Department of Physics and Astronomy, Tufts University,
Medford, MA}
\newcommand{\ucd}{Department of Physics, University of California, Davis, CA}
\newcommand{\unc}{Department of Physics, University of North Carolina, Chapel
Hill, NC}
\newcommand{\dresden}{Institut f\"{u}r Kern- und Teilchenphysik, Technische
Universit\"{a}t Dresden, Dresden, Germany}
\newcommand{\isu}{Department of Physics, Idaho State University, Pocatello,
ID}
\newcommand{\qmul}{Dept. of Physics, Queen Mary University, London, UK}
\newcommand{\ucsb}{Dept. of Physics, University of California, Santa Barbara,
CA}
\newcommand{\cern}{CERN, Geneva, Switzerland}
\newcommand{\utah}{Dept. of Physics, University of Utah, Salt Lake City, UT}

\author{
{B.~Aharmim}\altaffilmark{6}, 
{S.N.~Ahmed}\altaffilmark{14}, 
{A.E.~Anthony}\altaffilmark{17,a}, 
{N.~Barros}\altaffilmark{8}, 
{E.W.~Beier}\altaffilmark{13}, 
{A.~Bellerive}\altaffilmark{4}, 
{B.~Beltran}\altaffilmark{1}, 
{M.~Bergevin}\altaffilmark{7,5}, 
{S.D.~Biller}\altaffilmark{12}, 
{K.~Boudjemline}\altaffilmark{4}
% ...
}

\slugcomment{SNO Collaboration}

\altaffiltext{1}{\alta}
% ...
\altaffiltext{4}{\carleton}
\altaffiltext{5}{\uog}
\altaffiltext{6}{\lu}
\altaffiltext{7}{\lbnl}
\altaffiltext{8}{\lifep}
% ...
\altaffiltext{12}{\oxford}
\altaffiltext{13}{\penn}
\altaffiltext{14}{\queens}
% ...
\altaffiltext{17}{\uta}
% ...
\altaffiltext{a}{Present address: \casa} 
% ...
\altaffiltext{c}{Present address: \hu} 
% ...

\begin{abstract}
	We have performed three searches for high-frequency signals in the solar
neutrino flux measured by the Sudbury Neutrino Observatory (SNO), motivated by
the possibility that solar $g$-mode oscillations could affect the production or
propagation of solar $^8$B neutrinos.  The first search looked for any
significant peak in the frequency range 1/day to 144/day, with a sensitivity to
sinusoidal signals with amplitudes of 12\% or greater.  The second search
focused on regions in which $g$-mode signals have been claimed by experiments
aboard the SoHO satellite, and was sensitive to signals with amplitudes of 10\%
or greater.  The third search looked for extra power across the entire
frequency band.  No statistically significant signal was detected in any 
of the three searches.
\end{abstract}

\maketitle


\section{Introduction}
\label{sec:intro}

	Neutrinos are the only way known to directly probe the dynamics of the
solar core~\citep{Bahcall:1988}, and, through the Mikheev-Smirnov-Wolfenstein
(MSW) effect~\citep{Mikheev:1986, Wolfenstein:1977}, they can even carry information about the rest of
the solar envelope.  To date, however, converting measurements of solar
neutrino fluxes into constraints on solar models has proven to be
difficult~\citep{nussm}, because of the large number of
co-varying parameters upon which such models are built.  

	A relatively simple signal that could tell us something new about the
Sun would be a time-variation in the neutrino fluxes.  Over the past forty
years, measurements made by solar neutrino experiments have therefore been the
focus of many studies, ranging from attempted correlations with the solar
sunspot cycle to open searches for signals with periods of weeks or
months~\citep{sturrock:03, sturrock:04, sturrock:05, sk:periodicity, sno:periodicity}. 
The shortest period examined to date is
roughly one day, where the MSW effect predicts that neutrinos propagating
through the Earth's core during the night will undergo flavor transformation in
much the same way they do in the Sun, resulting in a net gain in the flux of
electron neutrinos ($\nu_e$s).   Although there have been occasional claims of
signals on timescales similar to known variations in the solar magnetic field,
in all cases there have been conflicting measurements that show the signals to
be spurious or absent entirely.  

	We present in this article the results of a search in a new frequency
regime for solar time variations. Our focus has been on signals whose periods
range from 24 hours down to 10 minutes.  The motivation for such a high
frequency search is in part the expectation for solar helioseismological
variations on scales of order an hour or less, in particular solar `gravity
modes' ($g$-modes)~\citep{christensen}.  These $g$-modes are non-radial oscillations that are
predicted to be confined to the solar core, and thus could in principle affect
either neutrino production or neutrino propagation.  The neutrinos that SNO
detects, those from $^8$B decay within the Sun, are particularly well-suited
for our search because they are created very deep within the solar core and
because their propagation is known to be sensitive to variations in the solar
density profile through the MSW effect.

	The effects of $g$-modes on solar neutrino fluxes have been examined by
Bahcall and Kumar~\citep{bk:gmode}, who sought to determine whether $g$-mode
effects could explain the apparent solar neutrino deficit, finding that any
effect was far too small to account for the roughly 60\% discrepancy.  More 
recently, ~\citet{burgess} looked at ways in which a broad spectrum of $g$-modes could alter the
expectation for a solar neutrino spectral distortion caused by the MSW effect.
Nevertheless, there are at this time no explicit predictions as to 
whether $g$-modes or any other short-timescale variations could
lead to measurable solar neutrino flux variations.

\section{Sudbury Neutrino Observatory}
\label{sec:sno}

	SNO was an imaging Cherenkov detector using heavy water (D$_2$O) as
both the interaction and detection medium~\citep{nim}.  The SNO cavern is
located in Vale Inco's Creighton Mine, at $46^{\circ} 28^{'} 30^{''}$ N latitude,
$81^{\circ} 12^{'} 04^{''}$ W longitude.  The detector resided 1783 m below sea
level with an overburden of $5890 \pm 94$~m water equivalent, deep enough that
the rate of cosmic-ray muons passing through the entire active volume was just
three per hour.

	One thousand metric tons of heavy water were contained in a 12~m
diameter transparent acrylic vessel (AV).  Cherenkov light produced by neutrino
interactions and radioactive backgrounds was detected by an array of 9456
Hamamatsu model R1408 20-cm photomultiplier tubes (PMTs), supported by a
stainless steel geodesic sphere (the PMT support structure or PSUP).  Each PMT was 
surrounded by a light concentrator (a `reflector'), which increased the
light collection to nearly $55$\%.   Over seven kilotonnes (7000 kg) of
light water shielded the heavy water from external radioactive backgrounds:
1.7~ktonne between the acrylic vessel and the PMT support sphere, and 5.7~ktonne
between the PMT support sphere and the surrounding rock. The 5.7~ktonne of light
water outside the PMT support sphere were viewed by 91 outward-facing 20-cm 
PMTs that were used for identification of cosmic-ray muons.  

	The detector was equipped with a versatile calibration deployment
system which could place radioactive and optical sources over a large range of
the $x$-$z$ and $y$-$z$ planes in the AV.  In addition, periodic `spikes' of 
short-lived radioactivity (such as $^{222}$Rn) were added to both the light water 
and heavy water and distributed throughout their volumes to act as distributed 
calibration sources.

 SNO detected neutrinos through three different processes: 

\begin{center}
  \begin{tabular}{ll}
     $ \nu_x + e^- \rightarrow \nu_x + e^-$  & (ES)\\
     $\nu_e + d \rightarrow p + p + e^-$\hspace{0.5in} & (CC)\\
     $ \nu_x + d \rightarrow p + n + \nu_x$ & (NC)\\  \\
  \end{tabular}
 \end{center}

where $\nu_x$ represents $\nu_e$, $\nu_{\mu}$ or $\nu_{\tau}$.  For both 
the Elastic Scattering (ES) and Charged Current (CC)
reactions, the recoil electrons were observed directly by their production of
Cherenkov light.  For the Neutral Current (NC) reaction, the neutrons were not
seen directly, but were detected when they captured on another nucleus.  In SNO
Phase I (the ``D$_2$O phase''), the neutrons captured on the deuterons present
within the SNO heavy water.  The capture on deuterium releases a 6.25~MeV
$\gamma$ ray, and it is the Cherenkov light of the secondary Compton electrons
or $e^+e^-$ pairs which was detected.  In Phase II, (the ``salt phase''), 2
tonnes of NaCl were added to the heavy water, and the neutrons captured predominantly on
$^{35}$Cl nuclei.  Chlorine has a much larger capture cross section (resulting
in a higher detection efficiency) for the neutrons.  The capture on chlorine
also yields multiple $\gamma$s instead of the single $\gamma$ from the pure
D$_2$O phase, which aids in the identification of neutron events.

\begin{figure}
\begin{center}
\includegraphics[width=5.0in]{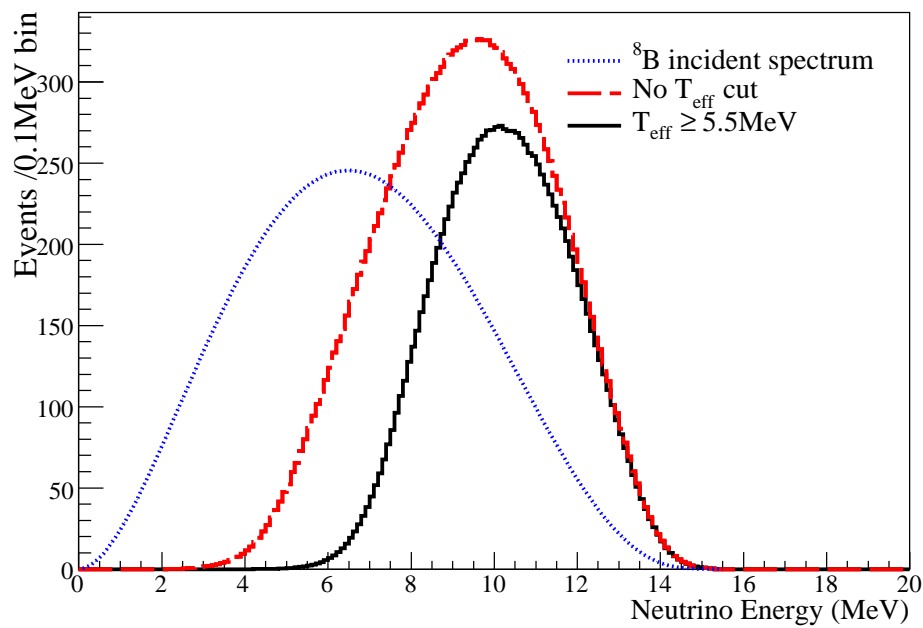}
\caption{Monte Carlo simulation of the $^8$B solar neutrino energy spectrum.  The dotted curve is the
incident spectrum of neutrinos from the Sun with an arbitrary normalization, the dashed curve shows the
spectrum of those
detected by the CC reaction before any cut on the kinetic energy of the created
electron, and the solid curve shows the spectrum of neutrinos seen after the
application of the kinetic energy threshold for Phase II. \label{fig:b8spec}}
\end{center}
\end{figure}

Figure~\ref{fig:b8spec} shows the incident $^8$B spectrum of neutrinos from the
Sun (dotted line), along with those that are detected by the CC reaction
(dashed) and those that are above the effective kinetic energy threshold for the
resultant electrons in  our detector for Phase II, $T_{\rm eff}>5.5$~MeV.  $T_{\rm eff}$  
is the estimated energy assuming an event consisted of a single electron.

	SNO's depth, very low radioactivity levels, and its ability to perform 
real-time detection made it a unique instrument for observing time
variations in solar neutrino fluxes, even at the high frequencies we examine
here.  Above an energy threshold of 5~MeV, the rate of events from
radioactivity, cosmic ray muons and atmospheric neutrinos, was negligible. The
rate of  solar neutrino events above this threshold was roughly 10/day.

\section{Data Sets}
\label{sec:datasets}

The event selection for the data sets is similar to that used in our lower-frequency periodicity 
analysis~\citep{sno:periodicity}.  Events were selected inside a
reconstructed fiducial volume of $R<550$~cm and above an effective
kinetic energy of $T_\text{eff}>5$~MeV (Phase I) or $T_\text{eff}>5.5$~MeV
(Phase II), and below $T_\text{eff}>20$~MeV.   Additional analysis cuts such as fiducial volume and background
rejection for these data sets have been described in detail
elsewhere~\citep{sno:longd2o,sno:nsp}.
SNO Phase I ran between November 2, 1999 and May 31, 2001, and with detector
deadtimes and periods of high radon removed, we recorded a total of 306.58 live
days, with 2924 candidate neutrino events.  SNO Phase II ran from July 26, 2001 to 
August 28, 2003 for a total of 391.71 live days and 4722 candidate neutrino events.  
Of the 2924 candidate events in the Phase I data, 67\% are due to
CC interactions, 20\% to NC interactions, and 9\% to ES
interactions, with the remaining 4\% due to backgrounds.
Variations in the $^8$B neutrino production rate itself will
affect all three neutrino interactions equally, while variations
in the electron neutrino survival probability dominantly affect
only the CC rate. For the Phase II data the 4722 events consist
of 45\% CC events, 42\% NC events, 6\% ES events, and 7\%
backgrounds.

The time for each event was measured with a global positioning system
(GPS) clock to a resolution of $\sim 100$~ns, but truncated to 10~ms accuracy for
the analysis.  The run boundary times were determined from the times of the
first and last events in each run with a precision of $\sim 50$~ms.
  The intervals between runs during which SNO was not recording solar neutrino
events correspond to run transitions, detector maintenance, calibration
activities, and any periods when the detector was off.  It is also necessary to
account for deadtime incurred within a run; for example, deadtime due to
spallation cuts that remove events occurring within 20 seconds after a muon.
This is important for a high-frequency periodicity search, as the frequency of
occurrence of these deadtimes can approach the scale of interest of our search.
Therefore, both the run boundaries, as well as the smaller, discrete breaks in
time due to removal of short-lived backgrounds such as spallation products,
define the time exposure of the data set, which itself may induce frequency
components that could affect a periodicity analysis.

\section{Rayleigh Power Approach}
\label{sec:rpwr}

	The low-frequency searches for periodicities that have been done by
ourselves and others typically group the neutrino time series in bins of one to
several days, and then perform the analysis with methods such as the
Lomb-Scargle technique~\citep{sk:periodicity, sturrock:04}.  In our own low-frequency
study~\citep{sno:periodicity}, we also used an unbinned maximum likelihood technique,
fitting the time series with periodic functions of varying frequencies, phases,
and amplitudes, allowing for the detector deadtimes that occurred during 
calibration runs, power outages, and detector maintenance.

	For this high-frequency study, we chose to use an unbinned 
Rayleigh power approach. The Rayleigh power of a time series for a given
frequency $\nu$ is defined as
\begin{equation}
  \label{eq:ray_form}
z(\nu) \equiv \frac{U(\nu)^2}{N} = \frac{1}{N} [ \left( {\Sigma}_i \cos{2\pi\nu t_i}\right)^2 + \left( {\Sigma}_i \sin{2\pi\nu t_i}\right)^2 ]
\end{equation}
where $N$ is the total number of events in the time series.

	The great advantage of the Rayleigh power approach is its speed, as it
requires far fewer function evaluations than other unbinned methods like the
maximum likelihood technique described above.  For this analysis, speed is
critical, because to ensure that we do not miss a signal we use 1.6 million equally-spaced 
frequencies spanning a range from 1/day to 144/day (one cycle per 10 minutes).  
To avoid the possibility of a signal falling between our sampled frequencies, and thus being missed, 
the minimum gap between our sampled frequencies must correspond to two signals
that just decorrelate over the course of SNO's running period.  With this
criterion, the minimum number of frequencies needed for our data set in our
region of interest is 400,000, and our choice of 1.6 million frequencies was
made to provide a small degree of oversampling.

	For a time series in which the phase coverage is uniform, the
distribution of Rayleigh powers for any given frequency follows $e^{-z}$, and
thus confidence intervals can be easily calculated.  For the SNO data set,
there are significant deadtime intervals, whose durations range from months to
milliseconds.  The sources of these deadtime intervals include the period
between the Phase I and Phase II data sets (several months), detector calibration
runs (typically hours to days), power outages (of order one day), maintenance
periods (hours), and offline veto periods (15~ms to 20~s) imposed to remove
events associated with the passage of cosmic-ray muons through the detector,
interactions of atmospheric neutrinos, or bursts of 
instrumental activity.  

	The deadtime structure of the SNO time series means that not all phases of
the Rayleigh power are equally likely, and thus leads to additional Rayleigh power
that is not associated with any neutrino signal.  Quasi-periodic deadtimes (like
those associated with calibration and maintenance) can also lead to peaks in the
Rayleigh power spectrum.  To calculate confidence intervals in order to determine
the significance of any peaks observed in the Rayleigh power spectrum, we must
account for these known regions of non-uniform phase coverage.

	We have developed an analytic model for the Rayleigh power at a given 
frequency by treating the Rayleigh power series as a two-dimensional random
walk. Each detected event is treated as a step in the random walk, with
components $X=\cos{2\pi\nu t}$ and $Y=\sin{2\pi\nu t}$.  For the case of
uniform phase coverage, the central limit theorem implies that for a large
number of steps  ($N$) the distribution of final 
positions will be given by a two-dimensional Gaussian, whose means, variances, and covariance are:
\begin{equation}
\mu_x = \frac{1}{2\pi} \int_0^{2\pi} d\phi~ \cos \phi = 0
\end{equation}
and
\begin{equation}
\sigma_x^2 = \frac{1}{2\pi} \int_0^{2\pi} d\phi~ \cos^2 \phi = \frac{1}{2}
\end{equation}
\begin{equation}
{\mathrm{cov}}(x,y) = 
\frac{1}{2\pi} \int_0^{2\pi} d\phi~ \cos \phi \sin \phi = 0
\end{equation}
leading to a simple distribution of final positions  given by
\begin{eqnarray*}
f(X,Y) & = & \frac{1}{2\pi N \sigma_x\sigma_y}
\exp(-X^2/2N\sigma_x^2) ~\exp(-Y^2/2N\sigma_y^2)  \\
 & = & \frac{1}{N\pi} e^{-(X^2+Y^2)/N} \\
 & = & \frac{1}{N\pi} e^{-U^2/N} \\
\end{eqnarray*}

	The distribution of $z=U^2/N$ can then be obtained by integrating over
all values of $X$ and $Y$ which satisfy $z<U^2/N < z+dz$, by changing variables
from $X$ and $Y$ to $\psi$ and $U$:
\begin{equation}
f(X,Y)~dX~dY \to f(U,\psi) ~dU~d\psi = \frac{1}{N\pi} e^{-U^2/N} ~U~dU~d\psi.
\end{equation}
Integrating this expression over $d\psi$ from 0 to $2\pi$ and changing variables
from $U$ to $z=U^2/N$ gives $f(z)~dz = e^{-z}~dz$, which is the simple exponential
distribution expected for the Rayleigh power distribution.

	For the case of non-uniform coverage, the means and covariance of the
distribution are no longer simple.  If we call the normalized phase-weighting function
$g(\phi)$, where $g(\phi)=1$ for uniform phase coverage, then we have
\begin{equation}
\label{eq:mux_gphi}
{\mu}_x = \frac{1}{2\pi} \int_0^{2\pi} d\phi~g(\phi) \cos \phi
\end{equation}

\begin{equation}
\label{eq:muy_gphi}
{\mu}_y = \frac{1}{2\pi} \int_0^{2\pi} d\phi~g(\phi) \sin \phi
\end{equation}

\begin{equation}
\label{eq:sigx_gphi}
{\sigma_x}^2 = \frac{1}{2\pi} \int_0^{2\pi} d\phi~g(\phi) ({\cos\phi - {\mu}_x})^2
\end{equation}

\begin{equation}
\label{eq:sigy_gphi}
{\sigma_y}^2 = \frac{1}{2\pi} \int_0^{2\pi} d\phi~g(\phi) ({\sin\phi - {\mu}_y})^2
\end{equation}

\begin{equation}
\label{eq:cov_gphi}
{\mathrm{cov}}(x,y) =\frac{1}{2\pi} \int_0^{2\pi} d\phi~g(\phi) (\cos \phi - {\mu}_x) (\sin
\phi - {\mu}_y)
\end{equation}

	The function $g(\phi)$ is determined by the detector's deadtime window, with
$\phi = \omega t$.  The mean for $X$, for example, is given by
\begin{equation}
\label{eq:mux_time}
{\mu}_x = \frac{1}{T} \displaystyle\sum_{j=1}^{runs} \int_{t_{start,j}}^{t_{stop,j}}
dt \cos \omega t
\end{equation}
	where $T$ is the total livetime, and the sum is over all data taking runs in
the data set, integrating from the start to the stop time of each run.  

	The Rayleigh power distribution for the non-uniform phase coverage case is
proportional to $e^{-\chi^2/2}$, and $\chi^2$ for the Rayleigh power distribution 
can be written as 
\begin{equation}
{\chi^{2}}\left(X,Y\right) = \left( X - N\mu_{x},Y - N\mu_{y}\right)
{V_{xy}}^{-1}
\left( \begin{array} {c}
X - N\mu_{x} \\
Y - N\mu_{y} \end{array} \right)
\end{equation}

The inverse of the covariance matrix is given by
\begin{center}
$ {V_{xy}}^{-1}={\left( \begin{array}{cc}
N{\sigma_{x}}^{2} & N{\mathrm{cov}}(x,y) \\
N{\mathrm{cov}}(x,y) & N{\sigma_{y}}^{2}\end{array} \right)}^{-1} $
\end{center}

making our $\chi^2$ 
\begin{equation}
{\chi}^{2} =
 \frac{{(X-N\mu_x)}^{2}{\sigma_y}^{2} + {(Y-N\mu_y)}^{2}{\sigma_x}^{2} -
2(X-N\mu_x)(Y-N\mu_y){\mathrm{cov}}(x,y)}
{N{\sigma_x}^{2}{\sigma_y}^{2}-N{\mathrm{cov}}^{2}(x,y)}
\end{equation}

Transforming into our integration variables $z=U^2/N$ and $\psi$ gives the rather 
unwieldy probability density function for the Rayleigh power at a given frequency: 
\begin{equation}
\label{eq:mess}
f(z) dz = \frac{1}{C} dz {\int}_0^{2\pi}{e^{-(\alpha_1(\psi)/2)Nz -
(\alpha_2(\psi)/2)N\sqrt{Nz} - (\alpha_3(\psi)/2)N^{2}}} \, d\psi
\end{equation}
where
\begin{eqnarray*}
&\alpha_1(\psi) = {\sigma_y}^2{\cos}^{2}\psi + {\sigma_x}^2{\sin}^{2}\psi  - 2
\cos\psi \sin\psi {\mathrm{cov}}(x,y) \\
&\alpha_2(\psi) = -2({\sigma_y}^2 \mu_x \cos\psi + {\sigma_x}^2 \mu_y \sin\psi   +
(\mu_x \sin\psi + \mu_y \cos\psi) {\mathrm{cov}}(x,y)) \\
&\alpha_3(\psi) = {\sigma_y}^2{\mu_x}^{2} + {\sigma_x}^2{\mu_y}^{2} + \mu_x
\mu_y {\mathrm{cov}}(x,y) 
\end{eqnarray*}
and $C$ is a normalization constant.   In this formalism the effects of
detector deadtime are entirely accounted for through the means and covariance
matrix for the variables X and Y.  In the analysis of data from the two combined 
SNO data-taking phases, the difference in event rates between Phase I and Phase II is 
accounted for by separating the terms of the analytic form according to phase, or: 
\begin{center}
$N \mu_{x,y} \rightarrow N_{D_2O} {(\mu_{x,y})}_{D_2O} + N_{Salt} {(\mu_{x,y})}_{Salt}$
\end{center}
(and similarly for all variance and covariance terms).  Here, both $g(\phi)$ and $N$ have been separated 
according to phase, effectively introducing a rate-dependent weighting factor.  For further 
details on this method, see Ref.~\citet{aea:thesis}.

\begin{figure}[!hp]
\begin{center}
\subfigure[Distribution of powers at sampled frequency = 1.000089 ${\rm day}^{-1}$.]
{
\includegraphics[width=2.55in]{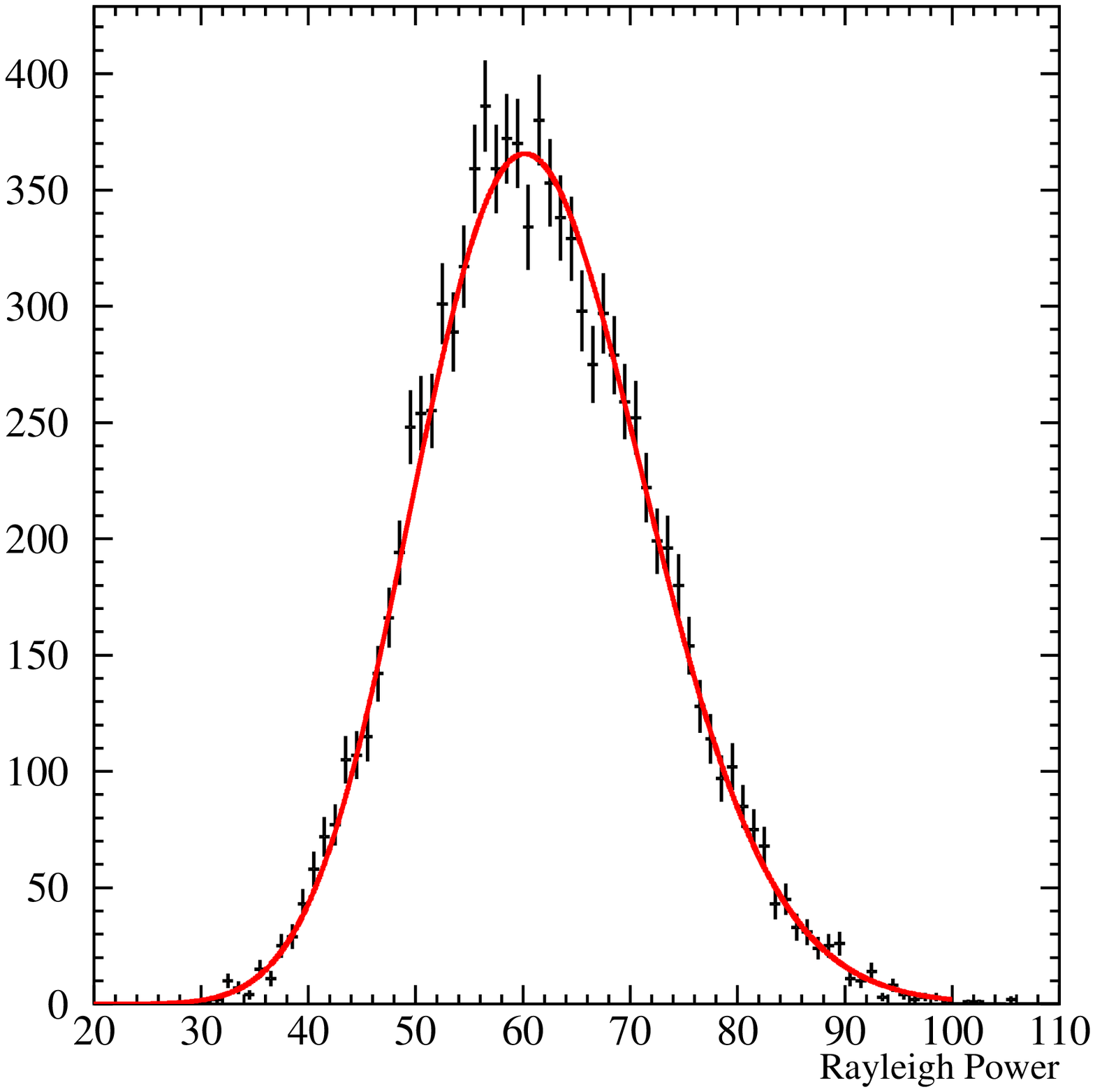}
}
\hspace{1cm}
\subfigure[Distribution of powers at sampled frequency = 1.046359 ${\rm day}^{-1}$.]
{
\includegraphics[width=2.55in]{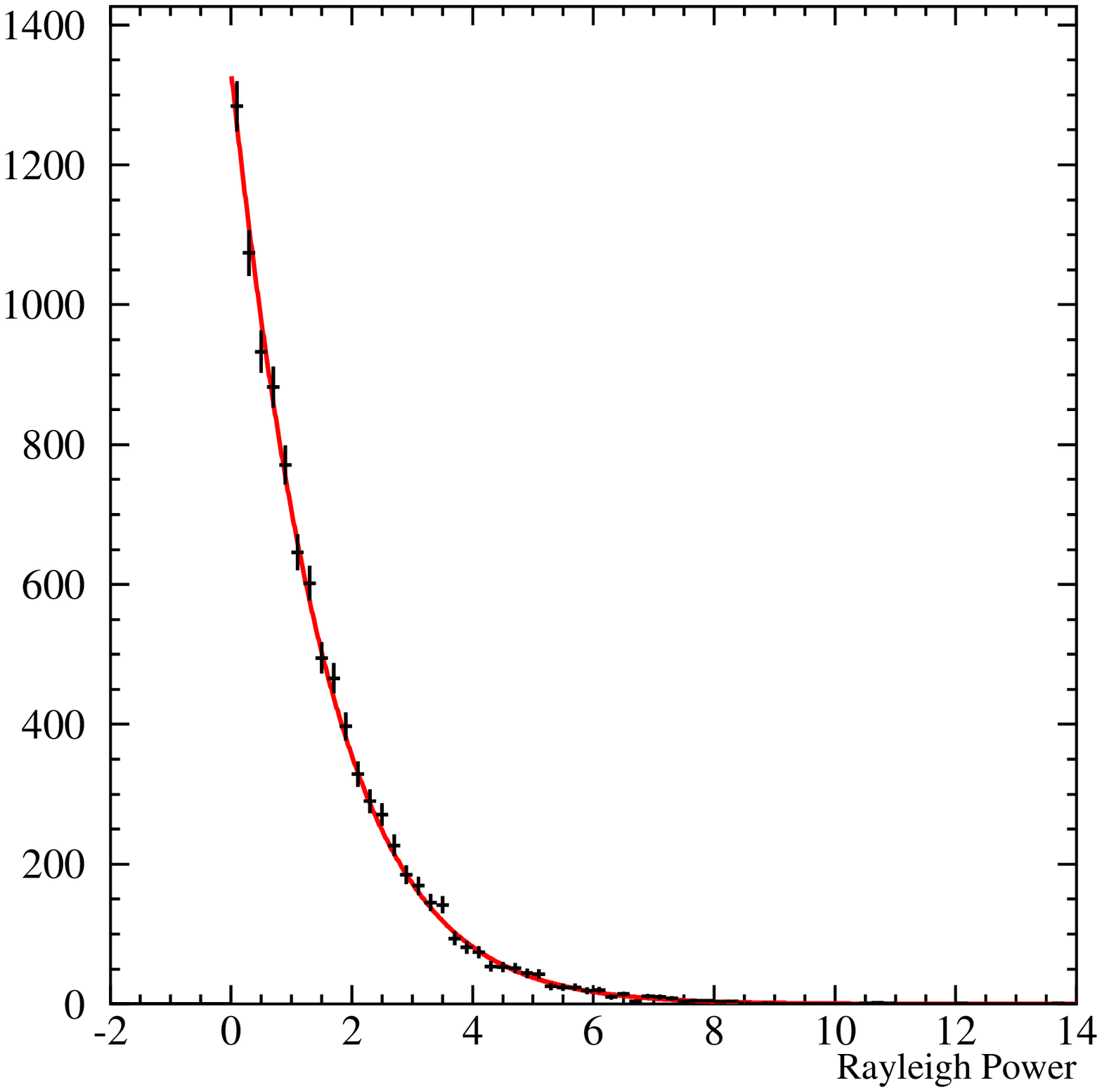}
}
\caption{Distribution of specific frequencies' Rayleigh powers in SNO Monte Carlo data sets (black) vs predictive analytic form (red line).\label{fig:fits}}
\end{center}
\end{figure}

	Figure~\ref{fig:fits} shows the function $f(z)$ describing the Rayleigh power
distributions for two different frequencies.  The distributions were generated
with a Monte Carlo simulation including the SNO detector's full deadtime window.
For the plot on the left, the deadtime contributes a significant amount of
power due to the periodicity in SNO's operations schedule,
while for the high frequency bin on the right the deadtime does not change the function much from its simple
$e^{-z}$ distribution.

To determine a specific confidence level, CL, for a given observed Rayleigh power, 
$z_0$, we solve the equation
\begin{equation}
{\rm CL} = \int_0^{z_0} f(z) dz
\label{eq:cl}
\end{equation}

\begin{figure}[!hp]
\begin{center}
\includegraphics[width=5.0in]{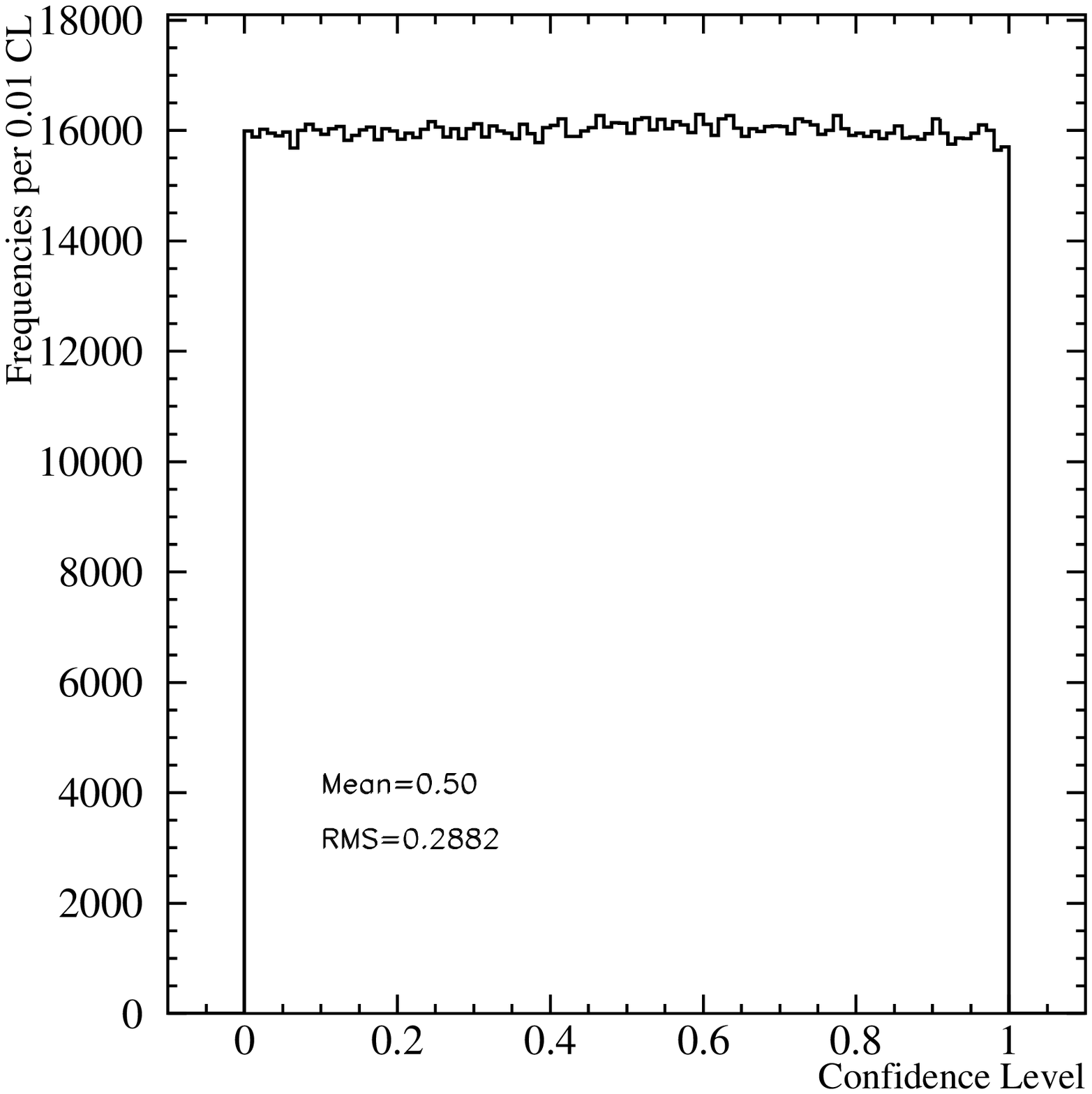}
\caption{Distribution of Rayleigh power confidence levels for all frequencies of a Monte Carlo simulation of the SNO  data set.\label{fig:clmctest}} 
\end{center}
\end{figure}

	As a test of our analytic model, we have calculated the confidence
levels for all 1.6 million frequencies of a Monte Carlo simulation that
includes the full SNO deadtime window.  Figure~\ref{fig:clmctest} shows the
distribution of these confidence levels, which is gratifyingly flat with a mean
that is 0.50, thus showing that the analytic random walk model correctly
distributes the confidence levels across the whole Rayleigh power spectrum.

\section{Open Single Peak Search}
\label{sec:open}

	Our first search looked for a significant peak at any frequency in our
Rayleigh power spectrum.  While Equation~\ref{eq:cl} gives the confidence level
at any specific frequency, when testing our 1.6 million sampled frequencies
there is a substantial trials penalty, making it exceedingly likely that at
least one of the frequencies will by chance have an apparently large power.  To
determine this penalty exactly, we would need to know how many of our 1.6
million sampled frequencies are independent, which is a complex task.

\begin{figure}[!hp]
\begin{center}
\includegraphics[width=4.5in]{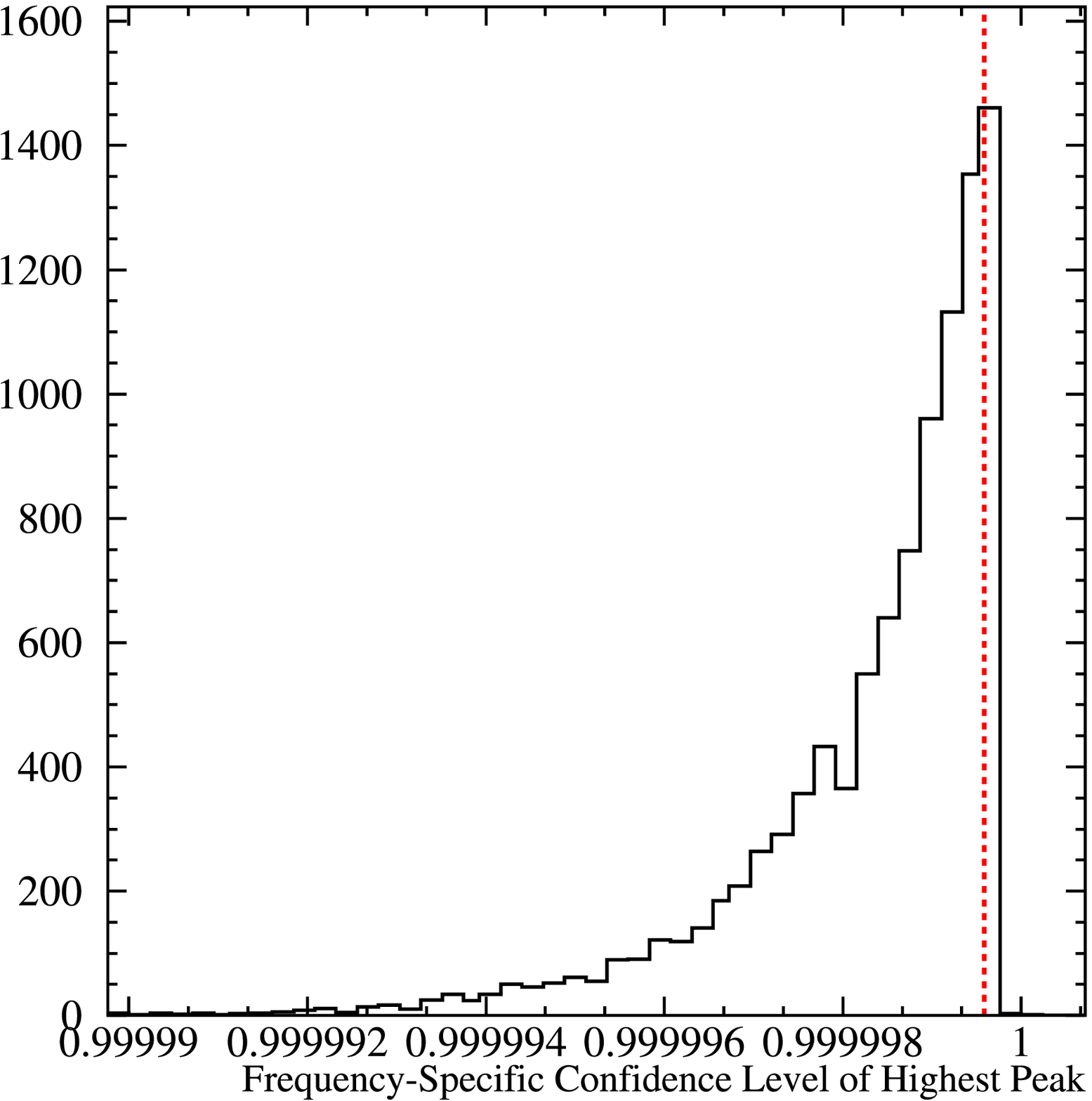}
\caption{Distribution of maximum confidence levels from Rayleigh analysis of 10,000 
signal-free Monte Carlo data sets.  By building this distribution of maximum confidence levels, we 
can determine a `confidence level of confidence levels' and account for the trials
penalty in our generation of the data-wide confidence level.  The frequency-specific 
confidence level corresponding to the data-wide confidence level of 90\% is shown 
with the superimposed dashed red line.\label{fig:cldist}}
\end{center}
\end{figure}

  Instead, to address this trials penalty we use 10,000 null-hypothesis Monte
Carlo simulations to determine the probability of observing a statistically
significant peak at any of the 1.6 million sampled frequency in the absence of
a true signal.  For a given null-hypothesis Monte Carlo simulation, we assign
a frequency-specific confidence level to each sampled frequency according to
the prescription given above in Section~\ref{sec:rpwr} (see Eq.~\ref{eq:cl}).
Then for each null-hypothesis Monte Carlo simulation, we record the peak that
has the highest confidence level, and then plot the resultant distribution of
these highest-peak confidence levels for all 10,000 null-hypothesis Monte Carlo
simulations. The resultant distribution of confidence levels is shown in
Figure~\ref{fig:cldist}. As seen from this figure, virtually every simulation
yields at least one peak with an apparent significance of at least 99.999\%,
just by chance.  To determine the true data-wide confidence level, taking
into account our entire sample of 1.6 million frequencies, we place a cut on the
distribution of Figure~\ref{fig:cldist} that corresponds to our desired
trials-weighted (data-wide, rather than frequency-specific) confidence level
for a significant signal. In Figure~\ref{fig:cldist}, the cut shown corresponds
to the frequency-specific confidence level needed by the maximum peak in a 
power spectrum in order for it to be above the 90\% CL detection threshold.  In other
words, for a data set that contains no periodicity at any frequency, there is a
less than 10\% probability that the most significant individual peak will have
a frequency-specific confidence level in excess of this cut value.

\begin{figure}[!hp]
\begin{center}
\includegraphics[width=4.0in]{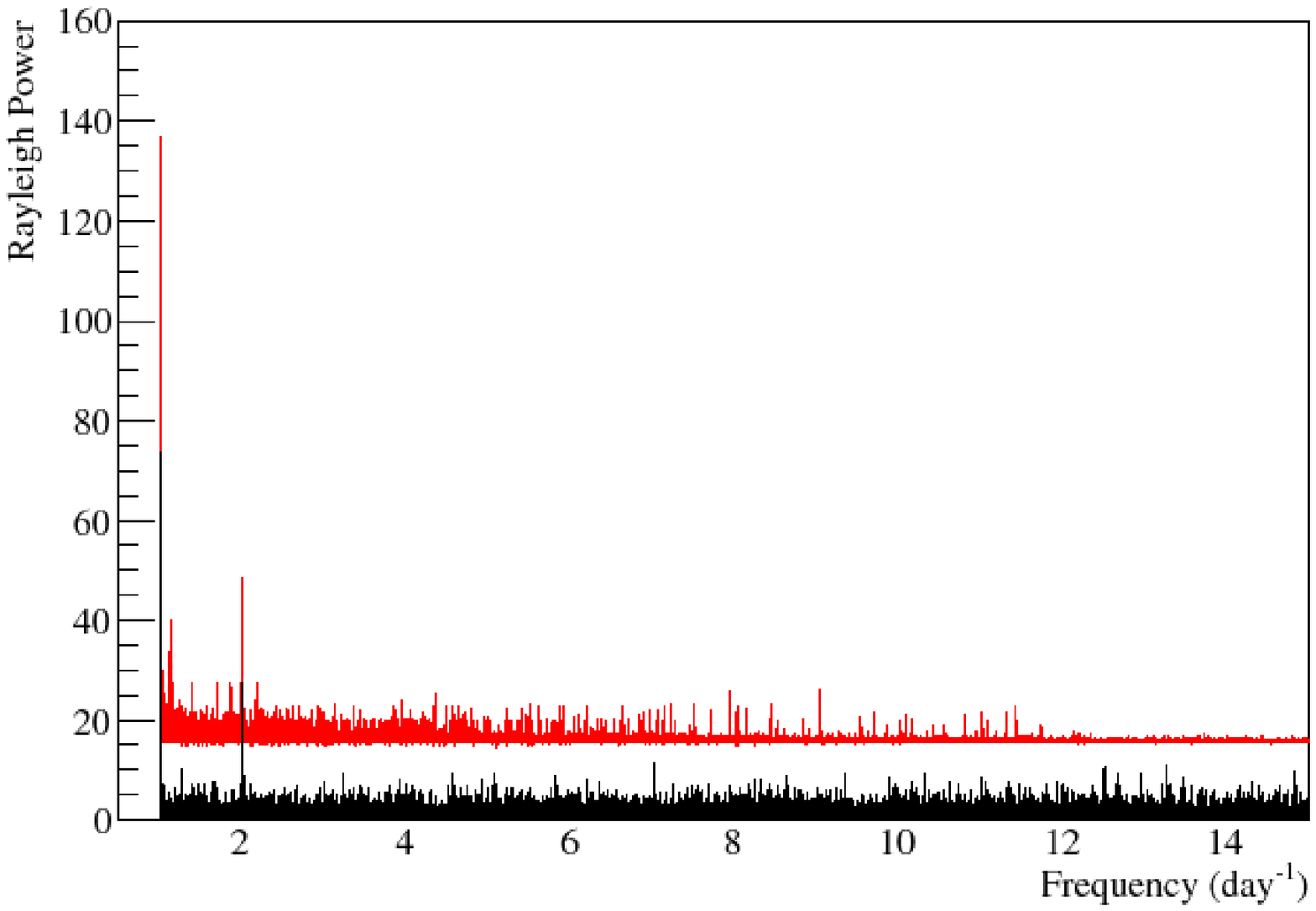}
\caption{First of six figures.  Rayleigh power spectra for first subsection (frequencies between 1 and 15 ${\rm day}^{-1}$) of entire 
range of frequencies sampled (from 1 ${\rm day}^{-1}$ to 144 ${\rm day}^{-1}$), for combined SNO Phase I and Phase II data sets.
The entire range has been broken down into six individual frames for easier inspection, with the first frame slightly more 
zoomed-in due to the presence of more underlying activity in this region of lower frequencies.  The black line indicates data, and 
the upper red line designates the level at which a detection would have a confidence level of 90\%.  The peaks at low frequencies, 
specifically those at 1${\rm day}^{-1}$ and 2${\rm day}^{-1}$, represent SNO-specific periodicities due to daily run-taking schedules.  
(This is clearly not evidence of a signal, as the red CL=90\% line, generated from null-hypothesis Monte Carlo, also follows these peaks).\label{fig:rpwr_frames_a}}
\end{center}
\end{figure}

\begin{figure}[!hp]
\begin{center}
\includegraphics[width=4.0in]{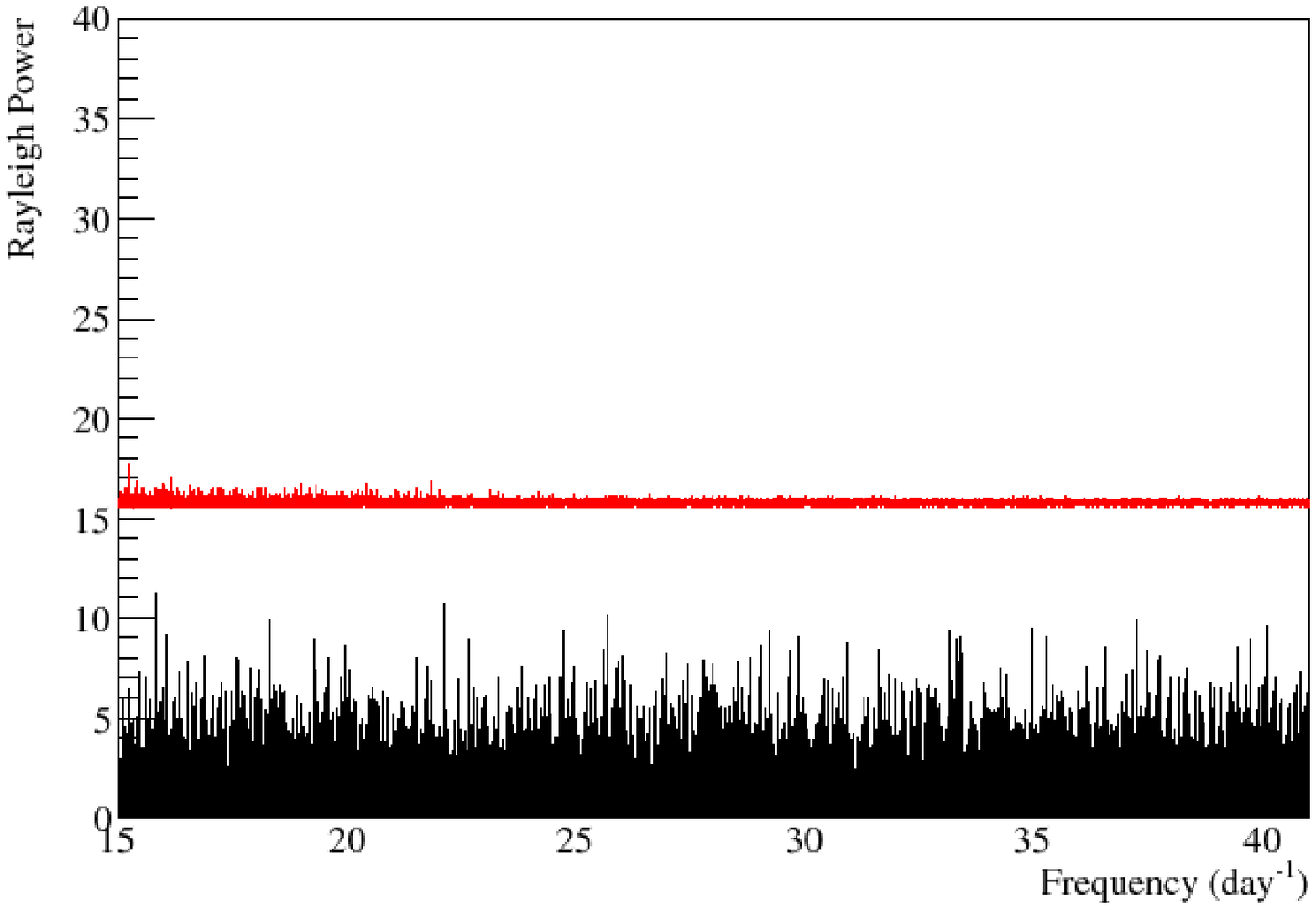}
\caption{Second of six figures.  Rayleigh power spectra for second subsection (frequencies between 15 and 41 ${\rm day}^{-1}$) of entire 
range of frequencies sampled (from 1 ${\rm day}^{-1}$ to 144 ${\rm day}^{-1}$), for combined SNO Phase I and Phase II data sets.  The black line 
indicates data, and the upper red line designates the level at which a detection would have a confidence level of 90\%.\label{fig:rpwr_frames_b}}
\end{center}
\end{figure}

\begin{figure}[!hp]
\begin{center}
\includegraphics[width=3.5in]{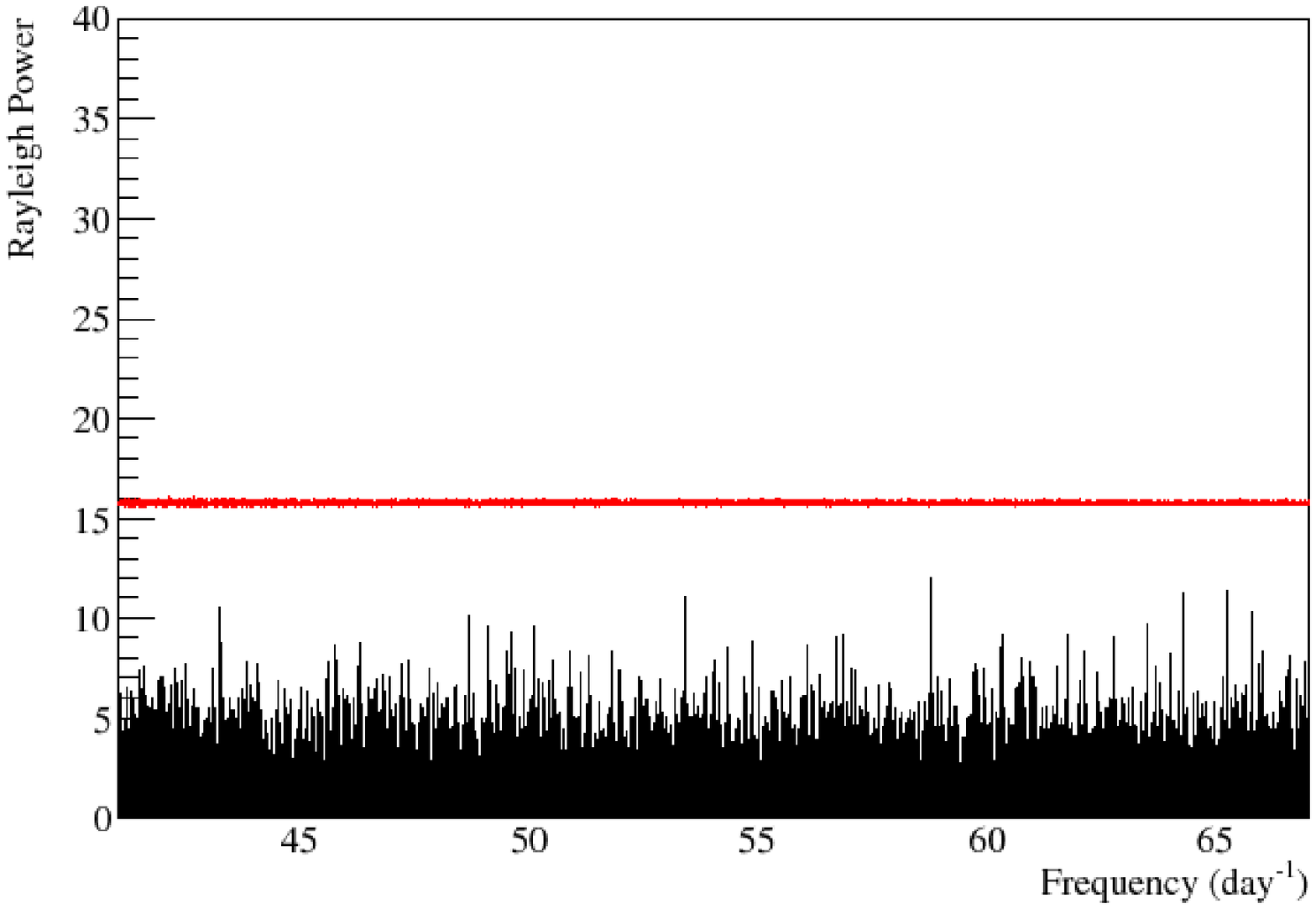}
\caption{Third of six figures.  Rayleigh power spectra for third subsection (frequencies between 41 and 67 ${\rm day}^{-1}$) of entire 
range of frequencies sampled (from 1 ${\rm day}^{-1}$ to 144 ${\rm day}^{-1}$), for combined SNO Phase I and Phase II data sets.  The black line 
indicates data, and the upper red line designates the level at which a detection would have a confidence level of 90\%.\label{fig:rpwr_frames_c}}
\end{center}
\end{figure}

\begin{figure}[!hp]
\begin{center}
\includegraphics[width=3.5in]{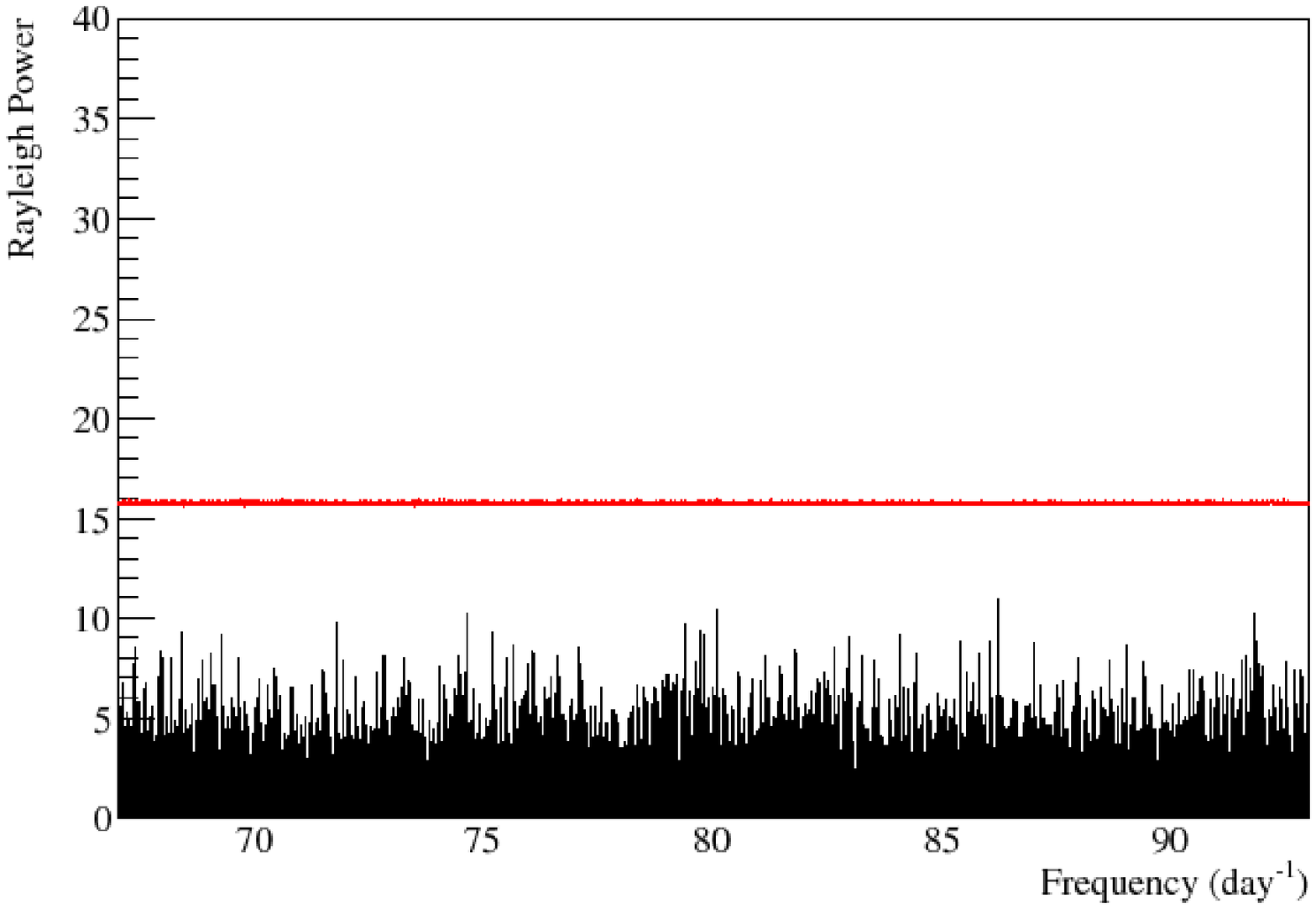}
\caption{Fourth of six figures.  Rayleigh power spectra for fourth subsection (frequencies between 67 and 93 ${\rm day}^{-1}$) of entire 
range of frequencies sampled (from 1 ${\rm day}^{-1}$ to 144 ${\rm day}^{-1}$), for combined SNO Phase I and Phase II data sets.  The black line 
indicates data, and the upper red line designates the level at which a detection would have a confidence level of 90\%.\label{fig:rpwr_frames_d}}
\end{center}
\end{figure}

\begin{figure}[!hp]
\begin{center}
\includegraphics[width=3.5in]{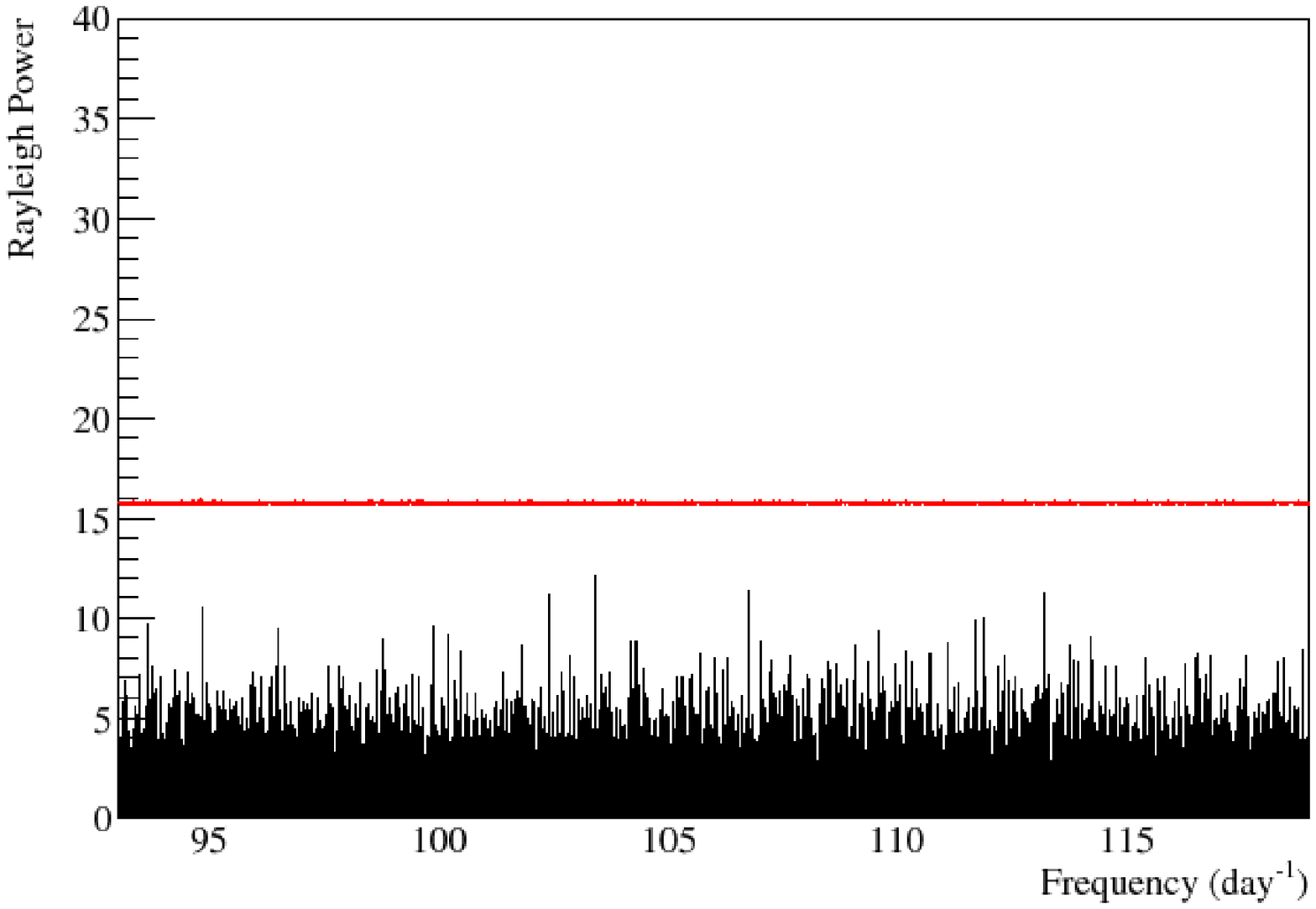}
\caption{Fifth of six figures.  Rayleigh power spectra for fifth subsection (frequencies between 93 and 119 ${\rm day}^{-1}$) of entire 
range of frequencies sampled (from 1 ${\rm day}^{-1}$ to 144 ${\rm day}^{-1}$), for combined SNO Phase I and Phase II data sets.  The black line 
indicates data, and the upper red line designates the level at which a detection would have a confidence level of 90\%.\label{fig:rpwr_frames_e}}
\end{center}
\end{figure}

\begin{figure}[!hp]
\begin{center}
\includegraphics[width=3.5in]{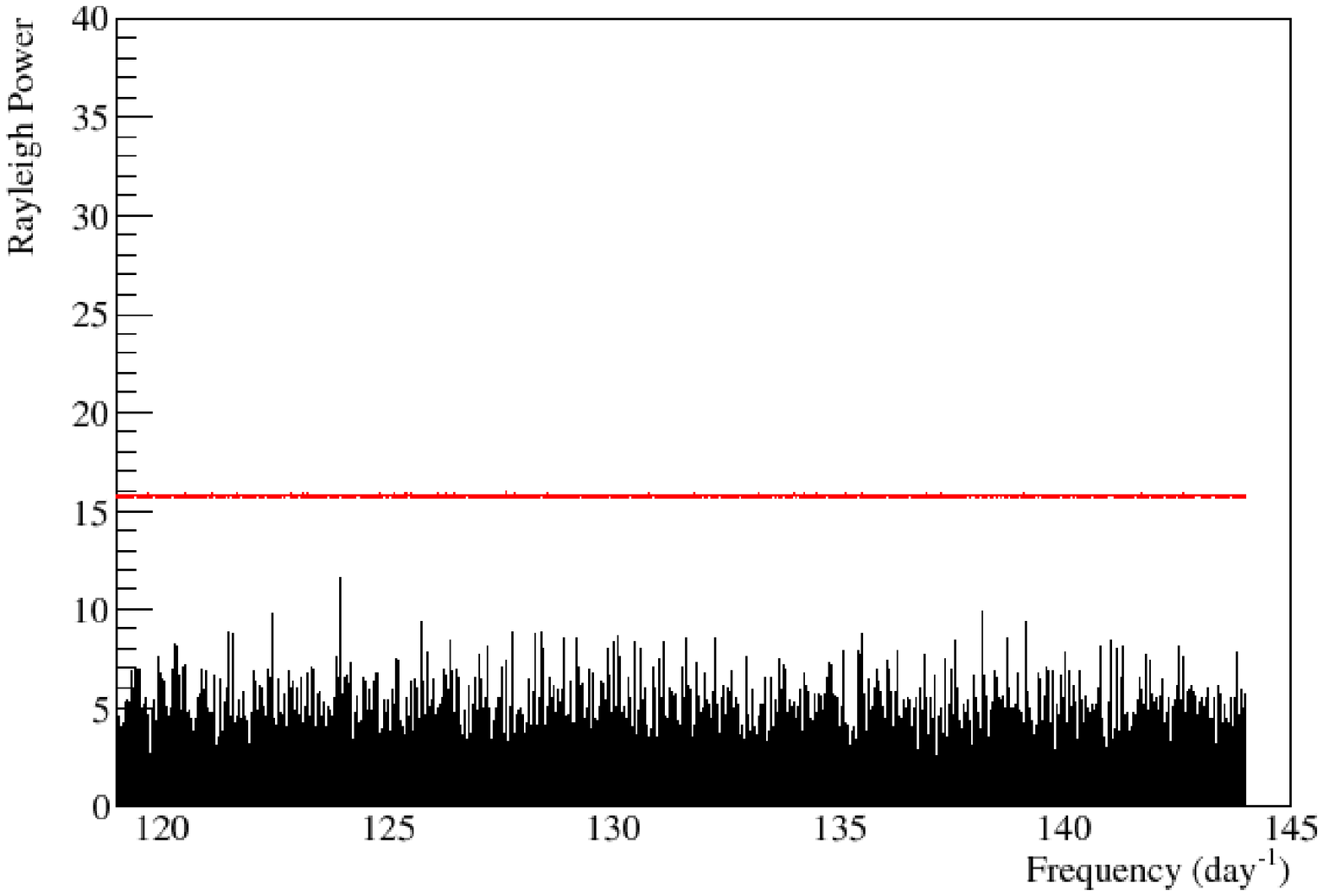}
\caption{Last of six figures.  Rayleigh power spectra for sixth subsection (frequencies between 119 and 144 ${\rm day}^{-1}$) of entire 
range of frequencies sampled (from 1 ${\rm day}^{-1}$ to 144 ${\rm day}^{-1}$), for combined SNO Phase I and Phase II data sets.  The black line  
indicates data, and the upper red line designates the level at which a detection would have a confidence level of 90\%.\label{fig:rpwr_frames_f}}
\end{center}
\end{figure}

	Figures~\ref{fig:rpwr_frames_a} through~\ref{fig:rpwr_frames_f} show the Rayleigh power spectrum for the
combined SNO Phase I and Phase II data sets, broken up into six segments each
corresponding to roughly 267,000 sampled frequencies.

\begin{figure}[!hp]
\begin{center}
\includegraphics[width=5.0in]{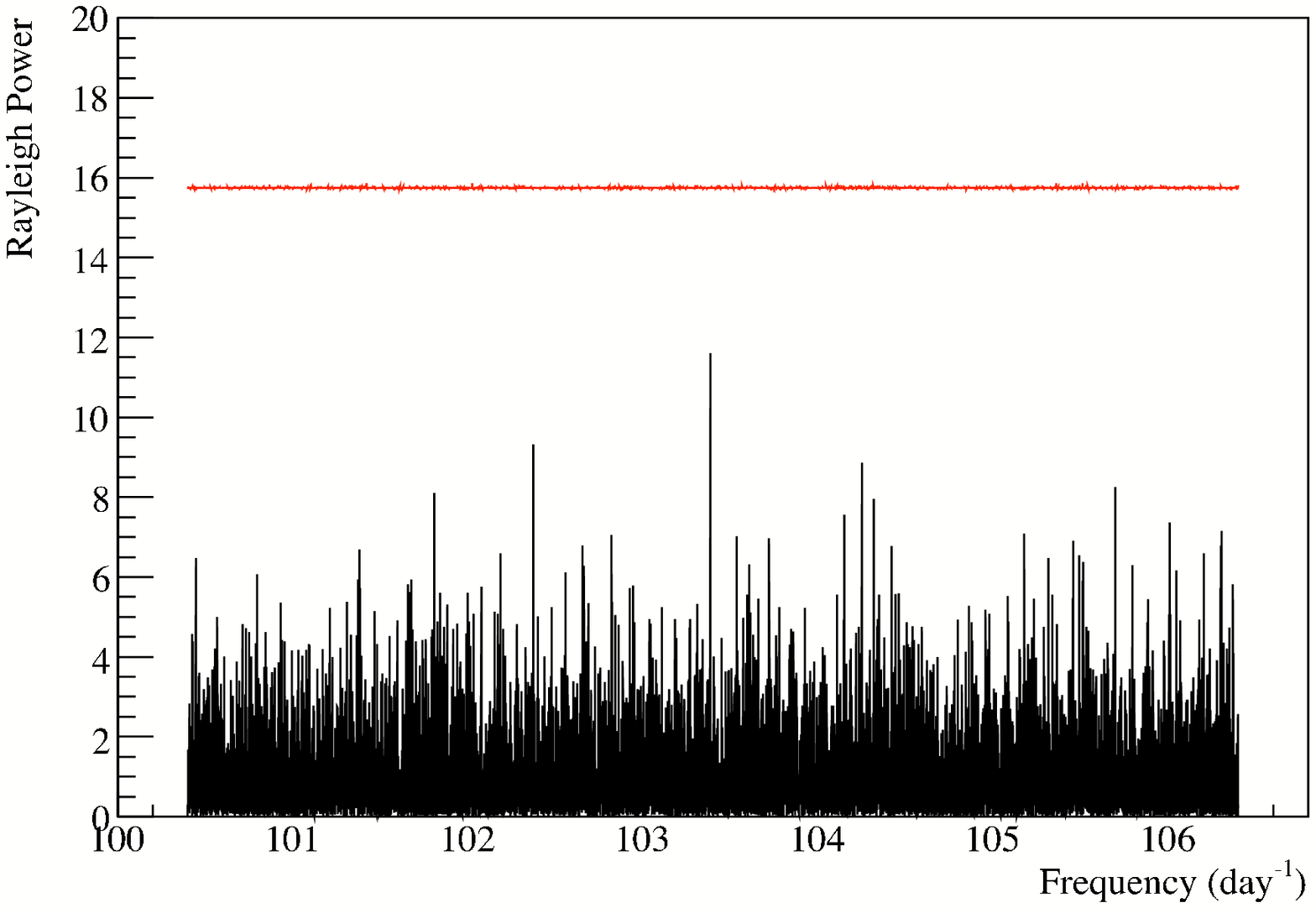}
\caption{Zoomed-in region of the Rayleigh power spectrum for the highest-significance peak in the 
SNO data set, which was detected at frequency=103.384 ${\mathrm day}^{-1}$, with 
a confidence level of 2\%. The horizontal red line indicates the
frequency-specific powers needed for a peak to be above the data-wide 90\% CL. \label{fig:maxcl}}
\end{center}
\end{figure}

  Figure~\ref{fig:maxcl} shows the peak
with the highest confidence level, and the corresponding threshold for that
peak to be above the data-wide 90\% CL to be considered significant.  The
data-wide CL of this peak is only 2\%, thus we see no evidence of a significant
peak in our data set.

\begin{figure} [!hp]
\begin{center}
\includegraphics[width=5.0in]{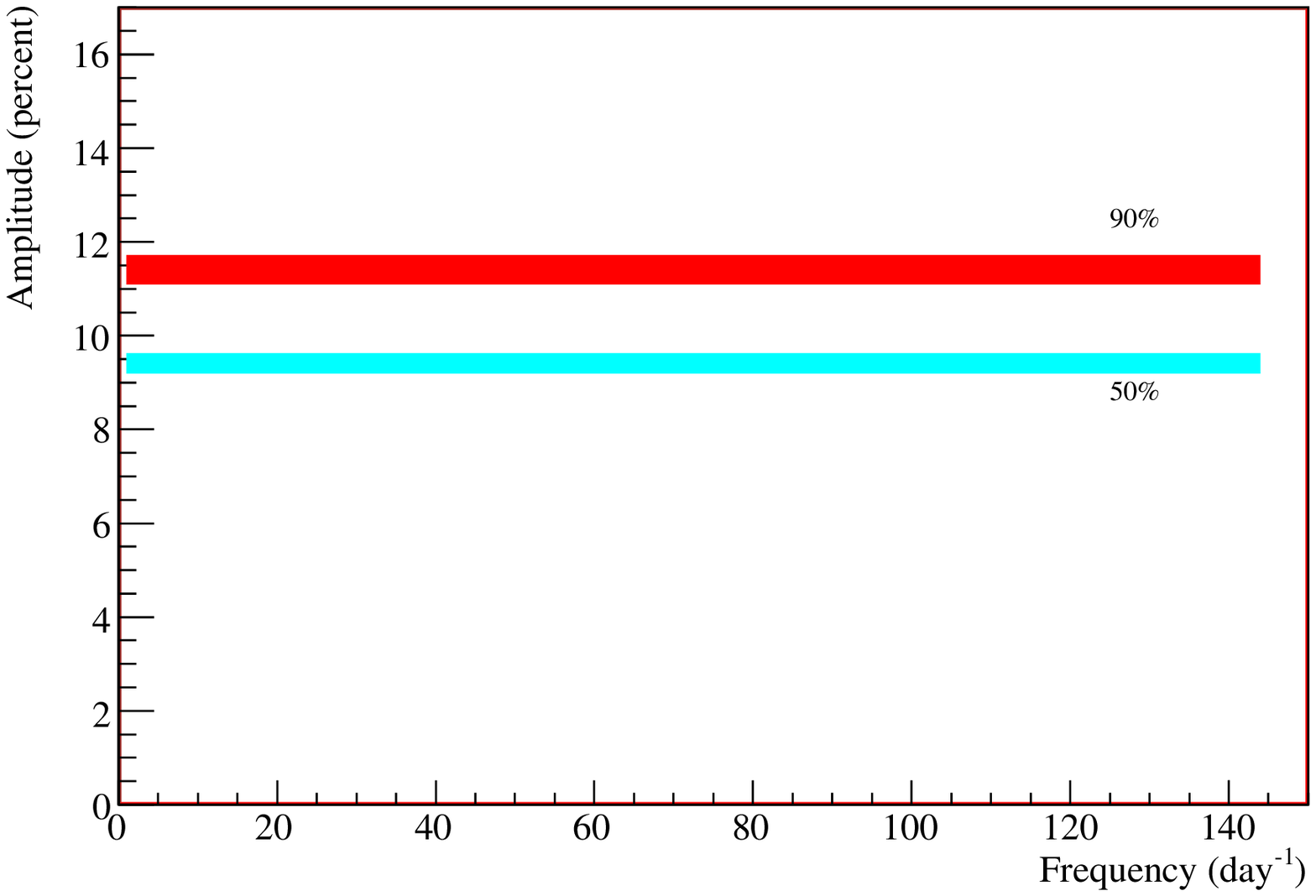}
\caption{SNO's sensitivities to a high-frequency periodic signal in the 
combined data sets (Phase I, or D$_2$O, and Phase II, or salt) for the entire frequency search region.  
The cyan band shows the calculated sensitivity at which we detect a signal 50\% of the time, with 99\% 
CL, and the red band shows the calculated sensitivity at which we detect a signal 90\% of the time, with 99\% 
CL.  The width of the bands represents the range of variation of the sensitivity, which varies rapidly with 
frequency, across the frequency regime.\label{fig:peaksens}}
\end{center}
\end{figure}

	To determine our sensitivity to a signal, we ran Monte Carlo
simulations with fake sinusoidal signals, of form $1 + A \sin \omega t$, of increasing amplitude, looking for the point
at which our method would claim a discovery.  In Figure~\ref{fig:peaksens} we
show our sensitivity for two criteria: the amplitude required to make a
99\%-CL discovery 90\% of the time, and the amplitude for a 99\%-CL 
discovery 50\% of the time.  We are substantially limited in this open search
by the trials penalty; we need a signal of 12\% amplitude to make a 99\% CL 
detection 90\% of the time.  The bands shown in Figure~\ref{fig:peaksens}
indicate the degree of variation among frequencies of the sensitivity, which is affected by the
underlying power spectrum in each bin as discussed above in
Section~\ref{sec:rpwr}.

\section{Directed Peak Search}
\label{sec:directed}

\begin{figure}[htp!]
\begin{center}
\includegraphics[width=5.0in]{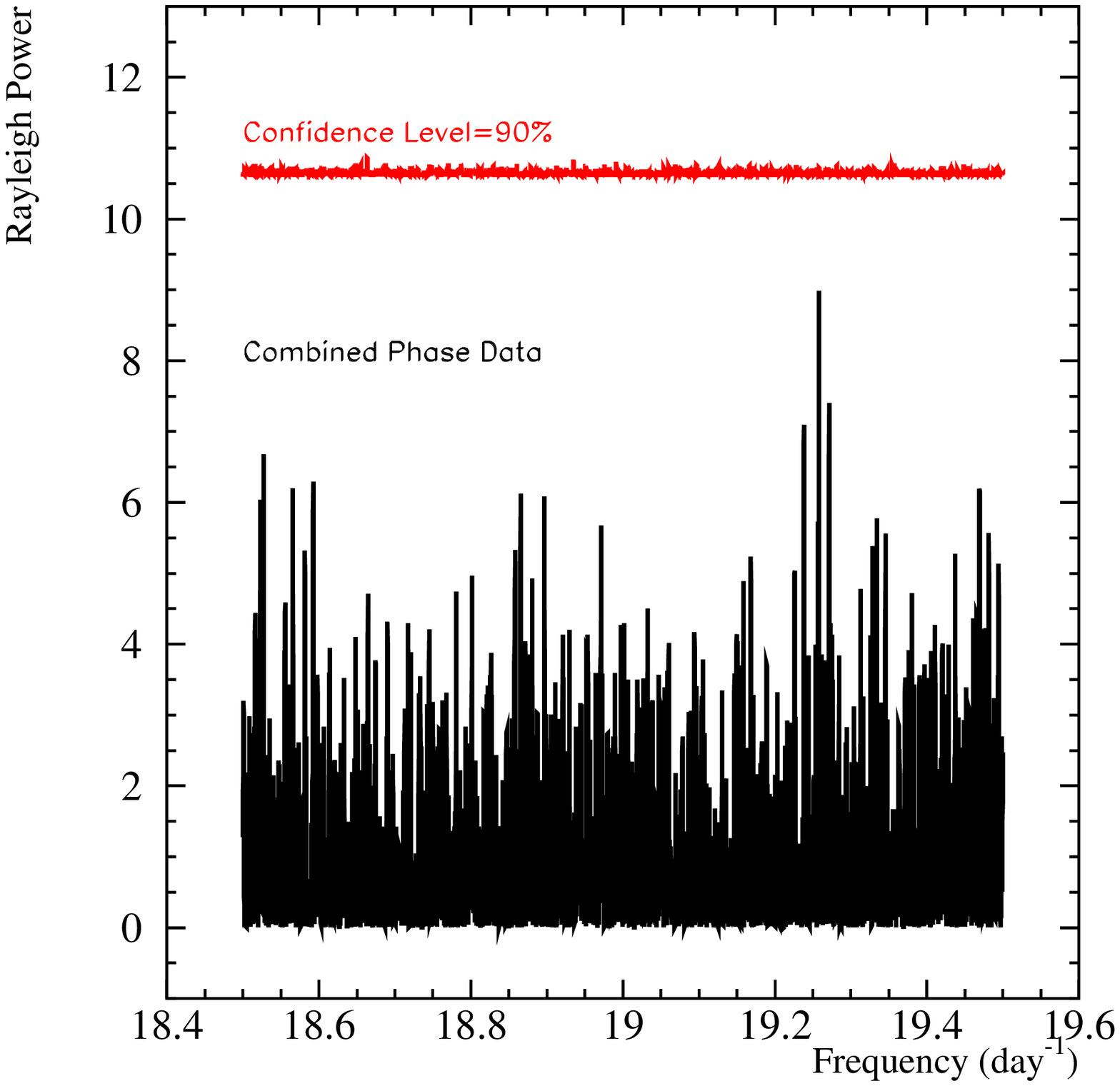}
\caption{Rayleigh power spectrum for `directed' high-frequency search, in black.  The line 
corresponding to detection with 90\% CL is shown in red.  The highest peak 
in the power spectrum is found at a frequency of 19.2579/day with a CL of 
58\%. \label{fig:dirpk}}
\end{center}
\end{figure}

\begin{figure}[htp!]
\begin{center}
\includegraphics[width=5.5in]{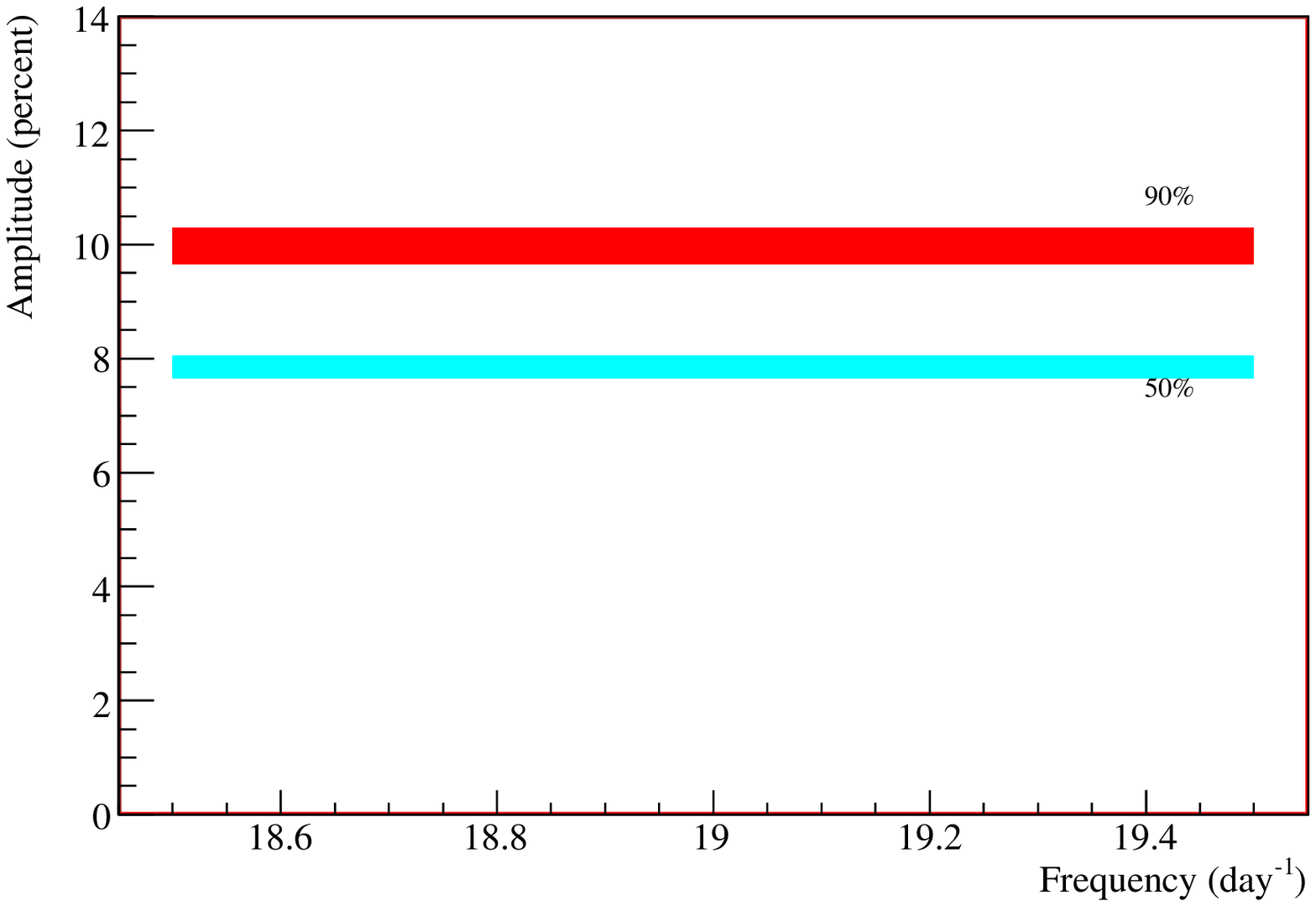}
\caption{SNO's sensitivities to a high-frequency periodic signal in the 
combined data sets (Phase I, or D$_2$O, and Phase II, or salt) for the directed high frequency search region.  
The cyan band shows the calculated sensitivity at which we detect a signal 50\% of the time, with 99\% 
CL, and the red band shows the calculated sensitivity at which we detect a signal 90\% of the time, with 99\% 
CL.  The width of the bands represents the range of variation of the sensitivity, which varies rapidly with 
frequency, across the frequency regime.\label{fig:dirsens}}
\end{center}
\end{figure}

	There have been recent claims by the GOLF/SoHO collaboration of
possible signatures of $g$-mode oscillations, based on analyses of long-term
helioseismological data sets~\citep{garcia01, gabriel02, turckchieze04, mathur07}, 
as well as supporting claims by the VIRGO/SoHO 
collaboration\citep{garcia08}.  Looking for such specific signals in our data
set using our Rayleigh power approach has an advantage in that we no longer
need the 1.6 million frequencies used above, but rather can look in a narrow
window that has a smaller trials penalty.  We have thus taken a narrow band
around the reported persistent GOLF signals~\citep{jimgarc08}, from 18.5 to 19.5/day 
(roughly 214 to 225 $\mu$Hz), 
and have repeated our analysis.  We again find no significant signal, with the 
highest peak having a trials-weighted CL of just 58\%.  The Rayleigh 
power spectrum around this peak, as well as its superimposed 90\% CL are 
shown in Figure~\ref{fig:dirpk}.  
Figure~\ref{fig:dirsens} shows our sensitivity plots for this directed search, 
which are slightly better than in Section~\ref{sec:open} because of the 
reduced trials penalty.  
We conclude that if the detection claimed by SoHO is in fact evidence of a \emph{g}-mode, the effect of this 
particular mode of oscillation on the neutrino flux is less than 10\% amplitude variation, at 99\% CL.

\section{Broadband Search}
\label{sec:broadband}

\begin{figure}[htp!]
\begin{center}
\includegraphics[width=5.0in]{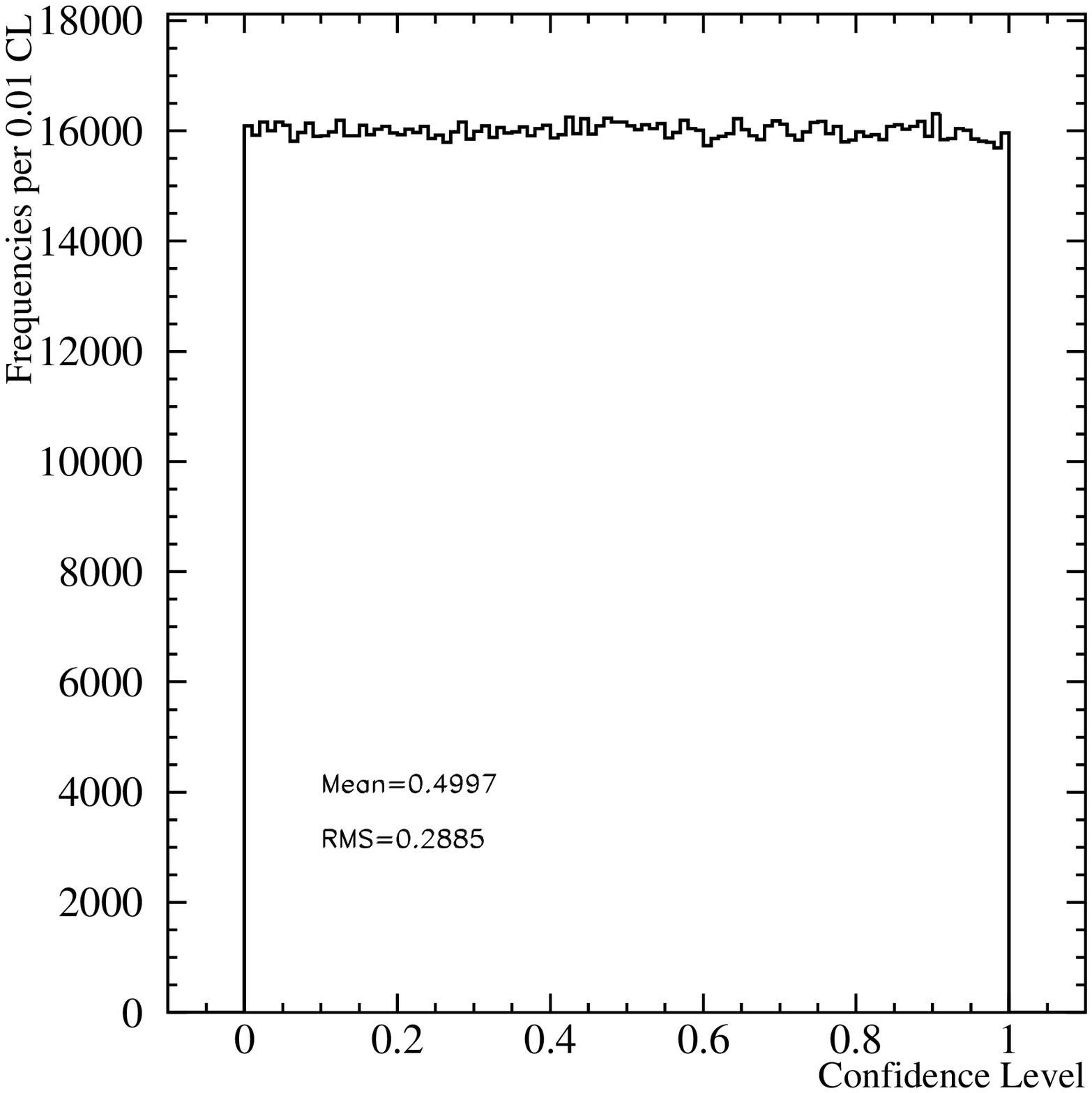}
\caption{Distribution of confidence levels for all 1.6 million frequencies, for the SNO combined-phase (D$_2$O and salt) data 
set.\label{fig:cldist_data}}
\end{center}
\end{figure}

\begin{figure}[!hp]
\begin{center}
\includegraphics[width=5.0in]{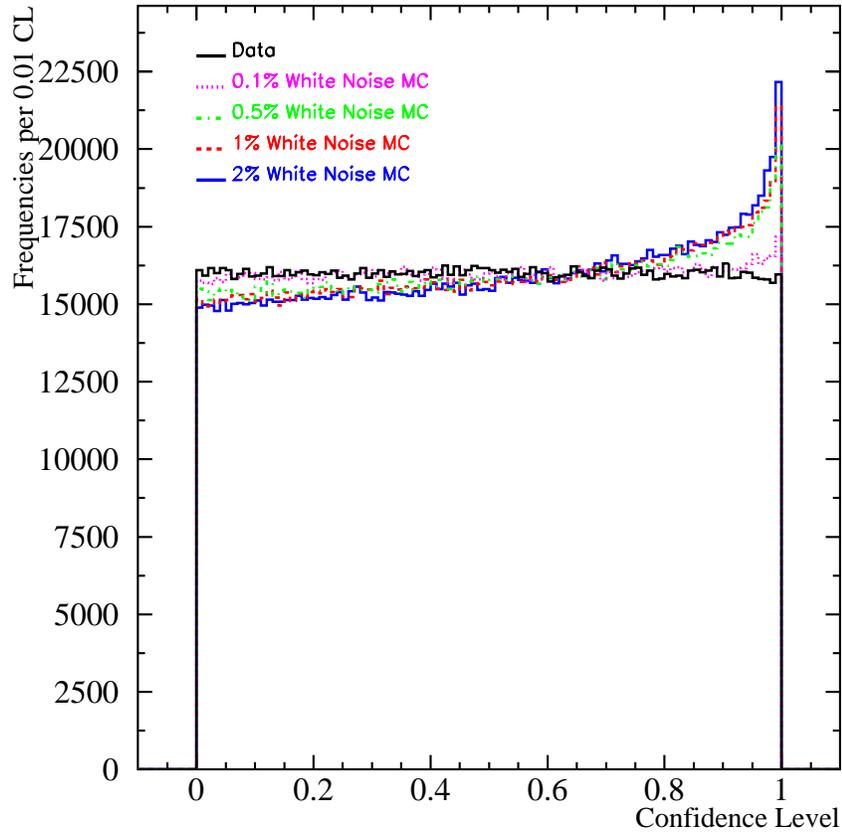}
\caption{Distribution of confidence levels for all 1.6 million frequencies in a SNO white-noise Monte Carlo, 
with several  signal amplitudes.  The SNO combined-phase (D$_2$O and salt) data confidence level distribution 
is shown in black for comparison. \label{fig:cldist_allnoise}}
\end{center}
\end{figure}

\begin{figure}[!hp]
\begin{center}
\includegraphics[width=5.0in]{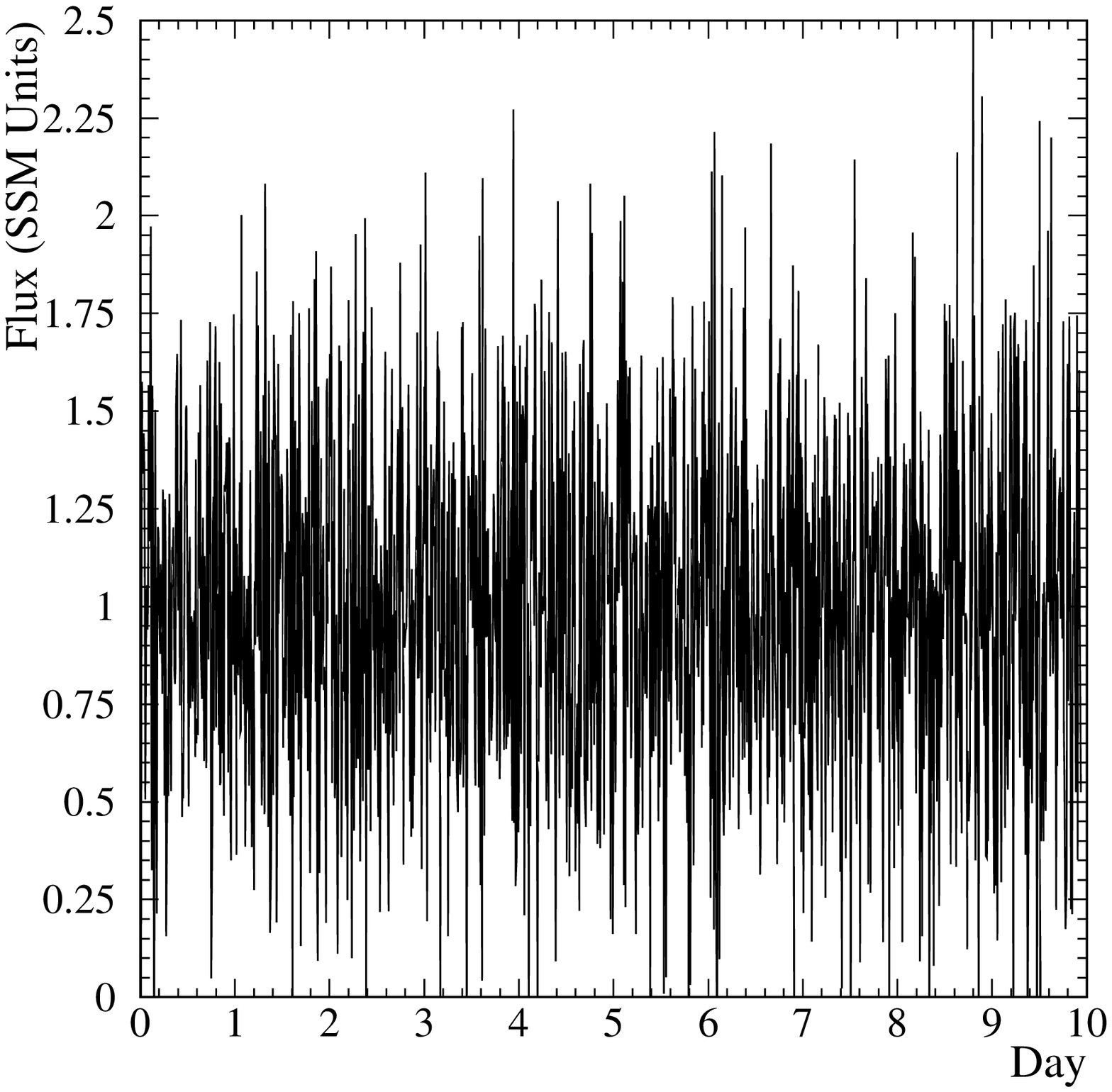}
\caption{Time domain plot of our Gaussian white noise model, with amplitude of
0.1\%, distributed over 400,000 frequencies.  Only ten days are shown here.
The rms noise power for the model shown here is 0.45 in units of SNO's measured
total $^8$B neutrino flux.
\label{fig:timenoise}}
\end{center}
\end{figure}

	The two searches described above require that the signal be
predominantly sinusoidal and monotonic.  It is possible that high frequency
behavior in the Sun spans a large band of frequencies, and in fact may be
`noisy'.  ~\citet{burgess} have investigated how such noise might affect
the neutrino survival probabilities within the Sun due to the matter or MSW
effect.  We have therefore looked at the distribution of confidence levels
across our entire range of 1.6 million frequencies.  Like Figure~\ref{fig:clmctest}, we
expect that in the case of no broadband signal the distribution of confidence
levels will be flat, with a mean of 0.50.  Figure~\ref{fig:cldist_data} shows
this distribution now for our combined SNO Phase I and Phase II data sets.  As can
be seen clearly in the figure, the distribution is flat, with a mean very close
to the 0.5 expected.  As a comparison case, we show in
Figure~\ref{fig:cldist_allnoise} what the confidence level distribution of a
`noisy' sun would look like, for several different amplitudes of Gaussian white
noise.  Our white noise model is shown in Figure~\ref{fig:timenoise}, for the
lowest amplitude (0.1\%) shown in Figure~\ref{fig:cldist_allnoise}.  We have
spread the noise across 400,000 independent frequencies, roughly the number of
independent frequencies we expect in the power spectrum.  The rms noise power
from our model is 0.45, in units of SNO's measured total $^8$B neutrino flux
($\sim 5\times 10^{6} \nu~{\rm cm}^{-2}{\rm s}^{-1}$).  In terms of power per unit 
bandwidth, this corresponds to $\sim 6\times 10^7 \nu~{\rm cm}^{-2}{\rm s}^{-1}{\rm Hz}^{-1/2}$.
As is evident in Figure~\ref{fig:cldist_data}, the distribution of confidence levels in the 
data is consistent with no distortion of signal due to noise.

\section{Conclusions}
	We have performed three searches for high-frequency signals in the
$^8$B solar neutrino flux, applying a Rayleigh Power technique to data from the
first two phases of the Sudbury Neutrino Observatory.  Our first search looked
for any significant peak in a Rayleigh Power spectrum from frequencies ranging
from 1/day to 144/day.  To account for SNO's deadtime window, we calculated the
expected distribution of power in each bin of the Rayleigh Power spectrum using
a random walk model, thus allowing us to assign confidence levels to the
observed powers.  We found no significant peaks in the data set. For this
`open' peak search, we had a 90\% probability of making a 99\% CL 
detection of a signal with an amplitude of 12\% or greater, relative to 
SNO's time-averaged neutrino flux.  

	In a second search, we narrowed our frequency band to focus on a region
in which $g$-mode signals have been claimed by experiments aboard the SoHO
satellite. The examined frequency range extended from 18.5/day to 19.5/day.  Again, no
significant peaks in the Rayleigh Power spectrum were found, and our
sensitivity for this `directed' search gave us a 90\% probability of making a 
99\% CL detection for signals whose amplitudes were 10\% or larger, relative
to SNO's time-averaged neutrino flux.

	Our third search examined the entire range of frequencies from 1/day to
144/day, looking for any evidence that additional power was present across the
entire high-frequency band.   To do this, we used the distribution of
frequency-specific confidence levels, determined using our random walk model.  We found
that, as expected for no high-frequency variations, this distribution was flat.
We showed that for a simple Gaussian white noise model, the confidence level
distribution would be noticeably distorted even when the amplitudes of the
contributing frequencies have an rms as small as 0.1\%.


\acknowledgments
This research was supported by: Canada: Natural Sciences and
Engineering Research Council, Industry Canada, National Research
Council, Northern Ontario Heritage Fund, Atomic Energy of Canada,
Ltd., Ontario Power Generation, High Performance Computing Virtual
Laboratory, Canada Foundation for Innovation; US: Dept. of Energy,
National Energy Research Scientific Computing Center; UK: 
Science and Technologies Facilities Council.  We thank the SNO 
technical staff for their strong contributions.  We thank Vale 
Inco, Ltd. for hosting this project.

\end{document}